\documentclass{article}

\usepackage{graphicx}
\usepackage{amssymb}
\usepackage{amsmath}
\usepackage{amsthm}
\usepackage{mathrsfs}
\usepackage{url}
\usepackage{color}
\usepackage{longtable}
\usepackage{rotating}
\usepackage{multirow}
\usepackage{float}
\usepackage{authblk}
\usepackage[framemethod=TikZ]{mdframed}
\usepackage{algorithm2e}
\usepackage{stmaryrd}

\definecolor{red}{rgb}{1,0,0}

\newtheorem*{mydef}{Definition}

\newfloat{Algorithm}{p}{Algorithm}[section]
\newtheorem{mythe}{Theorem}[section]

\begin{document}

\title{The Long Memory of Order Flow in the Foreign Exchange Spot Market}

\author{Martin D. Gould\footnote{Corresponding author. Email: m.gould@imperial.ac.uk.} \footnote{MDG completed part of this work while at the University of Oxford.} $^{\ddag}$, Mason A. Porter$^{\S \P}$, and Sam D. Howison$^{\S \|}$}

\affil{$\ddag$ CFM--Imperial Institute of Quantitative Finance, Imperial College, London SW7 2AZ\\$\S$ Oxford Centre for Industrial and Applied Mathematics, Mathematical Institute, University of Oxford, Oxford OX2 6GG\\$\P$ CABDyN Complexity Centre, University of Oxford, Oxford OX1 1HP\\$\|$ Oxford--Man Institute of Quantitative Finance, University of Oxford, Oxford, OX2 6ED}

\maketitle

\begin{abstract}We study the long memory of order flow for each of three liquid currency pairs on a large electronic trading platform in the foreign exchange (FX) spot market. Due to the extremely high levels of market activity on the platform, and in contrast to existing empirical studies of other markets, our data enables us to perform statistically stable estimation without needing to aggregate data from different trading days. We find strong evidence of long memory, with a Hurst exponent of $H \approx 0.7$, for each of the three currency pairs and on each trading day in our sample. We repeat our calculations using data that spans different trading days, and we find no significant differences in our results. We test and reject the hypothesis that the apparent long memory of order flow is an artifact caused by structural breaks, in favour of the alternative hypothesis of true long memory. We therefore conclude that the long memory of order flow in the FX spot market is a robust empirical property that persists across daily boundaries.\end{abstract}

\section{Introduction}

The autocorrelation properties of financial time series have been the subject of fierce debate for more than 50 years \cite{Cont:2005long,Cont:1997scaling,Gopikrishnan:1999scaling,Lo:1991long,Mandelbrot:1963variation}. Several important properties of financial markets have been reported to exhibit autocorrelations that decay slowly, often over periods of days or even months \cite{Booth:1979gold,Booth:1982rs,Chakraborti:2011empirical,Cont:1997scaling,Greene:1977long,Mantegna:1999introduction}. Such observations have prompted some authors to conjecture that some financial time series exhibit a phenomenon known as \emph{long memory} \cite{Baillie:1996long,Beran:1994statistics,Cont:2005long}, which means that the decay of autocorrelation is sufficiently slow that the sum of terms in their autocorrelation function (ACF) diverges to infinity.

In recent decades, the widespread uptake of electronic limit order books (LOBs) \cite{Gould:2013limit} in many financial markets has facilitated the recording of order-flow data, which provides a detailed description of traders' actions and interactions at the microscopic scale. The availability of such data has ignited interest in the possibility that financial markets exhibit long memory at the level of individual order flow, and several empirical studies during the past decade have reported this to be the case in a wide variety of different markets \cite{Bouchaud:2004fluctuations,Lillo:2004long,Mike:2008empirical,Toth:2015equity}.

In a recent publication, Axioglou and Skouras \cite{Axioglou:2011markets} challenged this view. They noted that, in order to construct sufficiently long time series to perform statistically stable estimation, existing studies of long memory in order flow have aggregated data from multiple trading days. Axioglou and Skouras argued that the apparent long memory reported by many studies is mostly an artifact caused by aggregating the data in this way. Specifically, they argued that the statistical properties of order flow change each day, and that concatenating order-flow series from different trading days creates nonstationarities at the boundaries between daily series. Moreover, statistical tests are known to produce similar output for nonstationary series as they do for stationary series with long memory \cite{Bhattacharya:1983hurst,Giraitis:2001testing,Granger:2004occasional}. Therefore, distinguishing between these alternatives is a difficult task.

Assessing whether or not order flow really is a long-memory process is important for several reasons. From a practical perspective, the present values of a long-memory process are correlated with values in the distant future \cite{Beran:1994statistics}, so identifying and quantifying the strength of long memory is useful for forecasting. From a theoretical perspective, several recent publications suggest that long-range autocorrelations in order flow may hold the key to understanding the complex statistical properties of price formation in financial markets (e.g., price impact, volatility, and the heavy-tailed distribution of returns \cite{Bouchaud:2009digest,Farmer:2006market,Gerig:2007theory,Toth:2011anomalous,Wyart:2008relation}). Moreover, if order flow really is a long-memory process, then identifying the sources of long-range autocorrelations may provide insight into traders' strategic decision-making processes \cite{Cont:2000herd,Toth:2015equity}.

In this paper, we perform an empirical study of a new, high-quality data set from a large electronic trading platform in the foreign exchange (FX) spot market to assess the long-memory properties of order flow for three liquid currency pairs. Due to the extremely high levels of market activity on the platform, and in contrast to existing empirical studies of other markets, our data enables us to perform statistically stable estimates of the long-memory properties of intra-day order flow without needing to aggregate data from different trading days. We are therefore able to exclude the possibility that our results are influenced by nonstationarities at the boundaries between different trading days, and we thereby avoid Axioglou and Skouras' \cite{Axioglou:2011markets} criticism of previous studies. For each of the three currency pairs and on all trading days in our sample, we find strong, statistically significant evidence for long memory in order flow.

To investigate how aggregating data from different trading days impacts our results, we also concatenate pairs of adjacent intra-day order-flow series to create cross-day series, which cross daily boundaries. We repeat all of our calculations on these cross-day series, and we find that our results are very similar to those for the intra-day series. We test and reject the hypothesis that the apparent long memory that we observe is an artifact caused by structural breaks, in favour of the alternative hypothesis of true long memory. We therefore conclude that the long memory of order flow in the FX spot market is a robust empirical property that persists across daily boundaries.

Several important differences separate our work from previous studies of long memory in order flow. First, our data originates from the FX market, whereas previous studies have analyzed data from equities markets \cite{Axioglou:2011markets,Bouchaud:2004fluctuations,Lillo:2004long,Mike:2008empirical,Toth:2015equity}. The FX market is the largest market in the world, so understanding its statistical properties is an important task. Second, the microstructural trade-matching rules in the FX market differ from those in equities markets. We provide a detailed comparison of these trading mechanisms in Section \ref{sec:data}. By examining how these differences impact traders' actions, we are able to gain additional insight into the underlying causes of the long memory that we observe. Third, because of the extremely high levels of activity in our data, we are able to perform statistically stable estimation of the long-memory properties of intra-day order flow without needing to aggregate data from multiple trading days. We are therefore able to test and reject Axioglou and Skouras' conjecture that the apparent long memory of order flow is mostly an artifact caused by aggregating data in this way. Fourth, we employ a wide range of estimators and statistical techniques to ensure that our quantitative assessment of long memory is robust. By contrast, some previous empirical studies have based their conclusions on single estimators, whose output can be misleading (as we demonstrate in Section \ref{subsec:lmresultsintra}). Together, these differences enable us to perform a detailed analysis of the strength, origins, and nature of long memory in order flow in a highly liquid but hitherto unexplored market.

The paper proceeds as follows. In Section \ref{sec:longmemoryinmarkets}, we provide a detailed discussion of long memory in order flow. In Section \ref{sec:litrev}, we review the findings of several empirical studies of long memory in order flow. We describe our data in Section \ref{sec:data}. We present our main results in Section \ref{sec:lmresults}, and we discuss our findings in Section \ref{sec:lmdiscussion}. We conclude in Section \ref{sec:lmconclusions}. In \ref{app:lmtests}, we present a technical overview of long memory. In \ref{sec:lmtests}, we discuss the statistical techniques that we use to assess the long-memory properties of order flow. In \ref{app:changepoint}, we discuss the statistical tools that we use to distinguish between a long-memory series and a short-memory series with nonstationarities.

\section{The Long Memory of Order Flow}\label{sec:longmemoryinmarkets}

In this paper, we perform an empirical analysis of the long-memory properties of order flow. First, we recall the formal definition of long memory and provide a detailed discussion of the order-flow series that form the basis of our empirical study.

\subsection{Long Memory}\label{subsec:longmemory}

Let\begin{equation}\label{eq:Wts}\left\{W_t\right\}=W_1,W_2,\ldots\end{equation}denote a real-valued, second-order stationary\footnote{A time series $\left\{W_t\right\}$ is \emph{second-order stationary} if its first and second moments are finite and do not vary with time \cite{Chatfield:2000time,Taylor:2008modelling}.} time series with mean\begin{equation}\label{eq:tsmean}\mathbb{E}(W_t)=\mu,\end{equation}autocovariance function\begin{equation}\label{eq:tsautocov}\gamma(k)=\mathrm{cov}\left(W_t,W_{t+\left|k\right|}\right),\end{equation}and autocorrelation function (ACF)\begin{equation}\label{eq:tsacf}\rho(k)=\frac{\gamma(k)}{\gamma(0)}.\end{equation}The time series $\left\{W_t\right\}$ is said to exhibit \emph{short memory} if\begin{equation}\label{eq:shortmemory}\lim_{N \rightarrow \infty}\sum_{k=-N}^N \left| \rho(k) \right|<\infty.\end{equation}The time series $\left\{W_t\right\}$ is said to exhibit \emph{long memory} if\begin{equation}\label{eq:longmemory}\lim_{N \rightarrow \infty}\sum_{k=-N}^N \left| \rho(k) \right|=\infty.\end{equation}For a technical introduction to long memory, see \ref{app:lmtests} and \cite{Beran:1994statistics}.

\subsection{Limit Order Books}\label{subsec:lobs}

More than half of the world's financial markets use limit order books (LOBs) to facilitate trade \cite{Rosu:2009dynamic}. In contrast to quote-driven systems, in which prices are set by designated market makers, trade in an LOB occurs via a continuous double-auction mechanism in which institutions submit orders. An \emph{order} $x=(p_x,\omega_x,t_x)$ submitted at time $t_x$ with price $p_x$ and size $\omega_x>0$ (respectively, $\omega_x<0$) is a commitment by its owner to sell (respectively, buy) up to $\left|\omega_x\right|$ units of the asset at a price no less than (respectively, no greater than) $p_x$.

Whenever an institution submits a buy (respectively, sell) order $x$, an LOB's trade-matching algorithm checks whether it is possible for $x$ to \emph{match to} an active sell (respectively, buy) order $y$ such that $p_y \leq p_x$ (respectively, $p_y \geq p_x$). If so, the matching occurs immediately and the owners of the relevant orders agree to trade the specified amount at the specified price. If not, then $x$ becomes \emph{active,} and it remains active until either it matches to an incoming sell (respectively, buy) order or it is \emph{cancelled}.

Orders that result in an immediate matching upon arrival are called \emph{market orders}. Orders that do not --- instead becoming active orders --- are called \emph{limit orders}.\footnote{Some platforms allow other order types (such as fill-or-kill, stop-loss, or peg orders \cite{HotspotGUIUserGuide}), but it is always possible to decompose the resulting order flow into limit and/or market orders. Therefore, we study LOBs in terms of these simple building blocks.} The \emph{LOB} $\mathcal{L}(t)$ is the set of all active orders for a given asset on a given platform at a given time $t$. For a detailed introduction to LOBs, see \cite{Gould:2013limit}.

Many LOBs record comprehensive digital transcriptions of order flow on a given platform. These transcriptions provide an event-by-event account of the temporal evolution of $\mathcal{L}(t)$, and they thereby enable detailed empirical analysis of financial markets at the microscopic scale \cite{Cont:2011statistical}.

\subsection{Order-Sign Series}\label{subsec:ordersignseries}

Given a sequence of $N$ consecutive arrivals of limit orders into $\mathcal{L}(t)$, the \emph{order-arrival series}\begin{equation}\label{eq:oas}\omega_{x_1},\omega_{x_2},\ldots,\omega_{x_N}\end{equation}is the time series of the arriving limit orders' sizes. Similarly, given a sequence of $M$ consecutive departures of active orders from $\mathcal{L}(t)$, the \emph{order-departure series}\begin{equation}\label{eq:ods}\omega_{x_1'},\omega_{x_2'},\ldots,\omega_{x_M'}\end{equation}is the time series of the departing active orders' sizes. An entry in the order-arrival series always corresponds to the arrival of a new limit order, but an entry in the order-departure series can occur either because an active order is cancelled or because an incoming market order triggers a matching and thereby removes a limit order from the LOB.\footnote{Incoming market orders are not reported in the order-arrival series.} Together, the order-arrival and order-departure series completely determine the temporal evolution of $\mathcal{L}(t)$.

When studying the long-memory properties of order-flow series, it is customary to study the time series of \emph{order signs}. For a given order $x_i$ of size $\omega_{x_i}$, the order sign $L_i$ is given by\begin{equation}\label{eq:loseries}L_i = \left\{\begin{array}{ll}
-1, & \text{if } \omega_{x_i}<0, \\
+1, & \text{if } \omega_{x_i}>0.\end{array}\right.\end{equation}Recall from Section \ref{subsec:lobs} that an order has negative size if and only if it is a buy order. Therefore, an order-sign series is simply a time series of $\pm 1$s, where $-1$ entries correspond to buy-order activity and $+1$ entries correspond to sell-order activity.

The reason for studying time series of order signs --- instead of the corresponding time series of order sizes --- is that empirical studies of a wide variety of different markets have reported that order sizes often vary over several orders of magnitude (see \cite{Gould:2013limit} for a recent survey of empirical studies of LOBs). This brings into question the convergence properties of higher-order moments of time series of order sizes. By contrast, studying only the time series of order signs guarantees that all moments exist, while still providing insight into the long-range autocorrelation properties of buy and sell activity in order flow.

\section{Literature Review}\label{sec:litrev}

Early studies of the autocorrelation properties of order flow tended to focus on short-range (i.e., small-lag) autocorrelations in order-sign series. Hasbrouck \cite{Hasbrouck:1988trades} studied the order-sign series for trades on the New York Stock Exchange (NYSE) during March--April 1985. He reported that lag-1 autocorrelations were strongly positive and that the mean sample ACF (see \ref{subsec:sampleacf}) across all stocks in the sample was positive up to lags of at least 200. Biais \emph{et al.} \cite{Biais:1995empirical} studied order-sign series for market orders, limit orders, and cancellations for 40 stocks traded on the Paris Bourse in 1991. For each type of order flow, they reported that any given event type (e.g., buy market order) was likely to be followed by another event of the same type. Ellul \emph{et al.} \cite{Ellul:2003comprehensive} and Yeo \cite{Yeo:2008serial} both reported similar findings for activity on the NYSE during 2001.

More recent work has focused on the long-memory properties of order-sign series. Lillo and Farmer \cite{Lillo:2004long} studied order-sign series for limit order arrivals, market order arrivals, and cancellations for 20 stocks on the London Stock Exchange (LSE) during 1999--2002. They used a wide variety of statistical techniques and estimators to test and reject the hypothesis that these series were short-memory series, in favour of the alternative hypothesis of long memory. They also estimated the Hurst exponent (see \ref{subsec:hurst}) for each series and reported a mean value of $H \approx 0.7$. The cross-sectional variation in $H$ across the stocks that they studied was small but significant. Because their sample ACFs contained no significant peaks or breaks corresponding to the length of a single trading day, Lillo and Farmer argued that long memory in order flow persists across daily boundaries. They also repeated their experiments on similar data from the NYSE and found similar results.

Bouchaud \emph{et al.} \cite{Bouchaud:2004fluctuations} studied long-range autocorrelations in the order-sign series for market orders on Euronext in 2001--2002. For all of the stocks that they studied, they reported that the sample ACFs decayed approximately according to a power law. They estimated each stock's power-law exponent directly from its sample ACF and reported values that correspond to Hurst exponents (see Equation (\ref{eq:hurstalpha})) ranging from $H \approx 0.65$ to $H \approx 0.9$. Similarly to Lillo and Farmer, Bouchaud \emph{et al.} argued that long-range autocorrelations in order flow persist across daily boundaries.

Mike and Farmer \cite{Mike:2008empirical} studied order-sign series for both market orders and limit orders for 25 stocks traded on the LSE during 2000--2002. They used detrended fluctuation analysis (DFA) (see \ref{subsec:lmdfa}) to estimate the Hurst exponent for each stock and reported values ranging from $H \approx 0.75$ to $H \approx 0.88$, with a mean of $H \approx 0.83$ across all stocks.

To date, two mechanisms have been proposed to explain the slow decay of autocorrelations in order-flow series. The first is that traders display ``herding'' behaviour, either because they all respond similarly to common information or because they monitor each other's actions and update their strategies by imitating those of their most successful competitors \cite{Lebaron:2007long}. The second is that traders who wish to perform large trades decompose them into smaller chunks, which they submit over several days (or even months) to minimize their market impact \cite{Bouchaud:2009digest,Bouchaud:2004fluctuations,Lillo:2005theory}. This strategy is commonly known as \emph{order splitting}.

Gerig \cite{Gerig:2007theory} assessed the plausibility of these two explanations by studying order-flow series from the LSE. In contrast to most LOB data sets, Gerig's data included information about the broker that submitted each order. This enabled him to compare the autocorrelation properties of order flow generated by individual brokers to those of the aggregate order flow generated by all brokers. He reported that correlations across different brokers decayed quickly to 0, but that autocorrelations in order flow from individual brokers exhibited long memory. He therefore argued that order splitting is a much more plausible explanation for long memory in order flow than is herding. T\'{o}th \emph{et al.} \cite{Toth:2012how,Toth:2015equity} also studied data containing brokerage identifiers from the LSE, and reached a similar conclusion.

Recently, however, Axioglou and Skouras \cite{Axioglou:2011markets} challenged the notion that order flow exhibits long memory by arguing that this apparent effect was mainly an artifact caused by nonstationarities in the underlying order-flow series. Specifically, they noted that to construct sufficiently long time series to perform statistically stable estimation of long-range autocorrelations, existing studies have aggregated order-flow data from different trading days. They argued that such aggregation produces order-flow series with structural breaks at the daily boundaries. Because many statistical tests produce similar output for nonstationary series as they do for stationary series with long memory (see \ref{app:changepoint}), Axioglou and Skouras conjectured the apparent long memory in order flow is mainly an artifact caused by these structural breaks.

To test this hypothesis, Axioglou and Skouras studied the order-sign series for market orders on the LSE during 2005--2006. They first aggregated data from several different trading days, and they noted that standard statistical tests applied to this data concluded strongly in favour of long memory. They then constructed shorter time series by aggregating data across pairs of consecutive trading days. They applied the cumulative-sum change-point estimator (see Equation (\ref{eq:berkesk})) to these series, and they were able to detect the daily boundaries with high accuracy. They then applied Berkes' change-point test (see \ref{app:changepoint}) to test the hypothesis that the apparent long memory in the cross-day series was actually due to a structural break. Working at the $5\%$ significance level, they could not reject the null hypothesis of a piecewise stationary series with a structural break in favour of the alternative hypothesis of true long memory in about two thirds of the order-flow series that they studied. They concluded that although order flow exhibited significant autocorrelations within a single trading day, the strength of autocorrelations that persisted across daily boundaries was very weak.

\section{Data}\label{sec:data}

We have been granted access to a recent, high-quality data set from Hotspot FX \cite{HotspotWebsite,HotspotGUIUserGuide}, which is one of the largest multi-institution trading platforms in the FX spot market. According to the 2010 Triennial Central Bank Survey \cite{BIS:2010triennial}, the mean daily turnover of the global FX market was approximately $4.0$ trillion US dollars. Approximately $37\%$ of this volume was due to spot trades, of which approximately $40\%$ was conducted electronically. In total, the mean daily volume traded on all multi-institution electronic trading platforms was approximately US $0.6$ trillion USD \cite{BIS:2010triennial}. The mean daily volume traded on Hotspot FX during the same period was approximately US $21.5$ billion USD \cite{HotspotVolumes}. Therefore, trade on Hotspot FX accounted for approximately $4\%$ of all volume traded electronically in the FX spot market during this period.

Hotspot FX offers trade for more than 60 different currency pairs. A price for the currency pair XXX/YYY denotes how many units of the \emph{counter currency} YYY are exchanged per unit of the \emph{base currency} XXX. Trade for each currency pair occurs in a separate LOB with price-time priority.

The Hotspot FX platform serves a broad range of trading professionals --- including banks, financial institutions, hedge funds, high-frequency traders, corporations, and commodity trading advisers \cite{HotspotWebsite}. As is customary on multi-institution trading platforms in the FX spot market, Hotspot FX enables institutions to specify \emph{credit limits} for their trading counterparties. Each institution can only access the trading opportunities offered by counterparties with whom they possess sufficient bilateral credit. We call this market organization a \emph{quasi-centralized limit order book (QCLOB),} because different institutions have access to different subsets of a centralized liquidity pool. For a detailed discussion of QCLOBs, see \cite{Gould:2015quasi}. Examples of platforms that utilize QCLOBs include Reuters \cite{ReutersWebsite}, EBS \cite{EBS:2011}, and Hotspot FX \cite{HotspotWebsite}. For more details about trade on Hotspot FX, see \cite{HotspotGUIUserGuide}.

The data that we study describes the full order-arrival and order-departure series (see Section \ref{subsec:ordersignseries}) for the EUR/USD (Euro/US dollar), GBP/USD (Pounds sterling/US dollar), and EUR/GBP (Euro/Pounds sterling) currency pairs during the \emph{peak trading hours} of $08$:$00$:$00$--$17$:$00$:$00$ GMT on 30 trading days during May--June 2010. Although trade in the FX spot market continues to operate outside of these hours, more than $70\%$ of the total traded volume for each of the three currency pairs occurs during these 9 hours each day.

For a given trading day $D_i$, we use the Hotspot FX data (see Section \ref{sec:data}) to produce an ordered list of the limit order arrivals that occur during the peak trading hours of $08$:$00$:$00$--$17$:$00$:$00$ GMT. We then use Equation (\ref{eq:loseries}) to deduce the intra-day arrival sign series from this list. Similarly, we use the Hotspot FX data to produce an ordered list of the active order departures that occur during the same period, and we then use Equation (\ref{eq:loseries}) to deduce the intra-day departure sign series from this list. We repeat this process for each of the $30$ trading days $D_1,D_2,\ldots,D_{30}$ in our sample.

In Table \ref{tab:sfxallagg}, we list the minimum, maximum, and mean number of limit order arrivals and departures across these 30 trading days. The number of arriving limit orders is largest for EUR/USD and smallest for EUR/GBP. For each of the three currency pairs and on most trading days that we study, the total number of limit order arrivals slightly exceeds the total number of departures, which implies that active orders accumulate throughout the trading day.

\begin{table}[!htbp]
\small
\begin{center}
\begin{tabular}{|c|c|c|c|c|}
\hline
 & & EUR/USD & GBP/USD & EUR/GBP \\
\hline
\multirow{3}{*}{Number of Arrivals} & Minimum & $3455561$ & $2962688$ & $2019826$ \\ 
& Maximum & $6003406$ & $5296372$ & $3623053$ \\ 
& Mean & $4533550.8$ & $4340345.4$ & $2932726.8$ \\ 
\hline
\multirow{3}{*}{Number of Departures} & Minimum & $3449793$ & $2961217$ & $2019672$ \\ 
& Maximum & $5992343$ & $5293082$ & $3622559$ \\ 
& Mean & $4524175.4$ & $4337320.6$ & $2932171.5$ \\ 
\hline
\end{tabular}
\caption{Minimum, maximum, and mean number of limit order (top panel) arrivals and (bottom panel) departures for EUR/USD, GBP/USD, and EUR/GBP, measured across the 30 trading days in our sample.}
\label{tab:sfxallagg}
\end{center}
\end{table}

To quantify the imbalance between buying and selling activity, we also calculate the percentage of sell orders (i.e., percentage of $+1$ entries in the order-sign series) among limit order arrivals and departures each day (see Table \ref{tab:sfxallimb}). For each of the three currency pairs and for both arrivals and departures, sell orders account for close to $50\%$ of order flow. Therefore, the level of buying activity is approximately equal to the level of selling activity on all trading days that we study.

\begin{table}[!htbp]
\small
\begin{center}
\begin{tabular}{|c|c|c|c|c|}
\hline
 & & EUR/USD & GBP/USD & EUR/GBP \\
\hline
\multirow{3}{*}{Percentage of Arriving Sell Orders} & Minimum & $48.664\%$ & $49.389\%$ & $48.817\%$ \\ 
& Maximum & $50.870\%$ & $51.255\%$ & $50.252\%$ \\
& Mean & $50.008\%$ & $50.135\%$ & $49.894\%$ \\
\hline
\multirow{3}{*}{Percentage of Departing Sell Orders} & Minimum & $48.667\%$ & $49.389\%$ & $48.815\%$ \\ 
& Maximum & $50.879\%$ & $51.257\%$ & $50.251\%$ \\
& Mean & $50.009\%$ & $50.134\%$ & $49.895\%$ \\
\hline
\end{tabular}
\caption{Minimum, maximum, and mean percentages of sell orders among limit order (top panel) arrivals and (bottom panel) departures for EUR/USD, GBP/USD, and EUR/GBP across the 30 trading days in our sample.}
\label{tab:sfxallimb}
\end{center}
\end{table}

\section{Results}\label{sec:lmresults}

In this section, we present our empirical results for the arrival-sign series. The corresponding results for the departure-sign series are qualitatively similar.

\subsection{Results for Intra-Day Series}\label{subsec:lmresultsintra}

\begin{figure}[!htbp]
\centering
\includegraphics[width=0.7\textwidth]{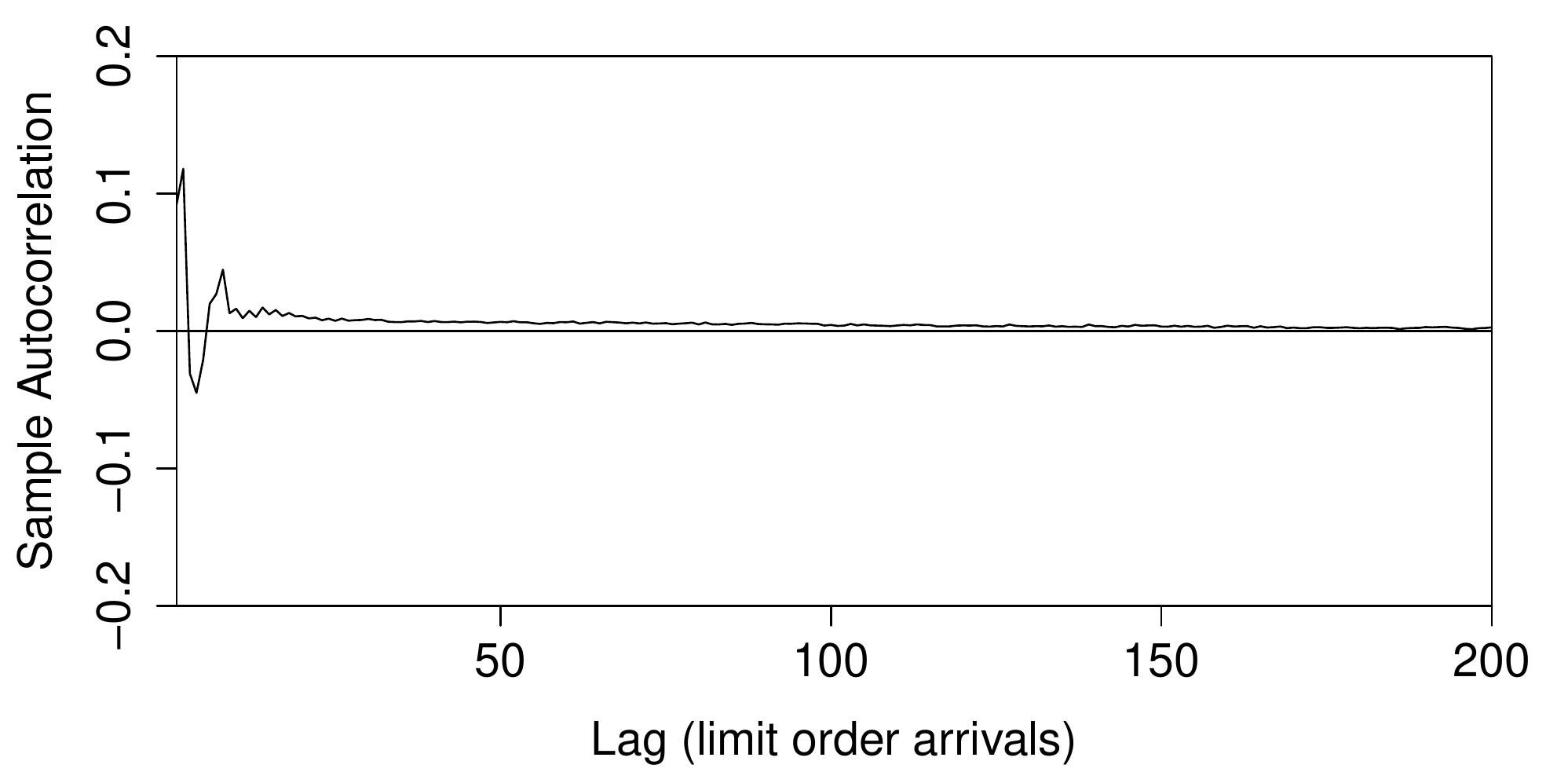}
\includegraphics[width=0.7\textwidth]{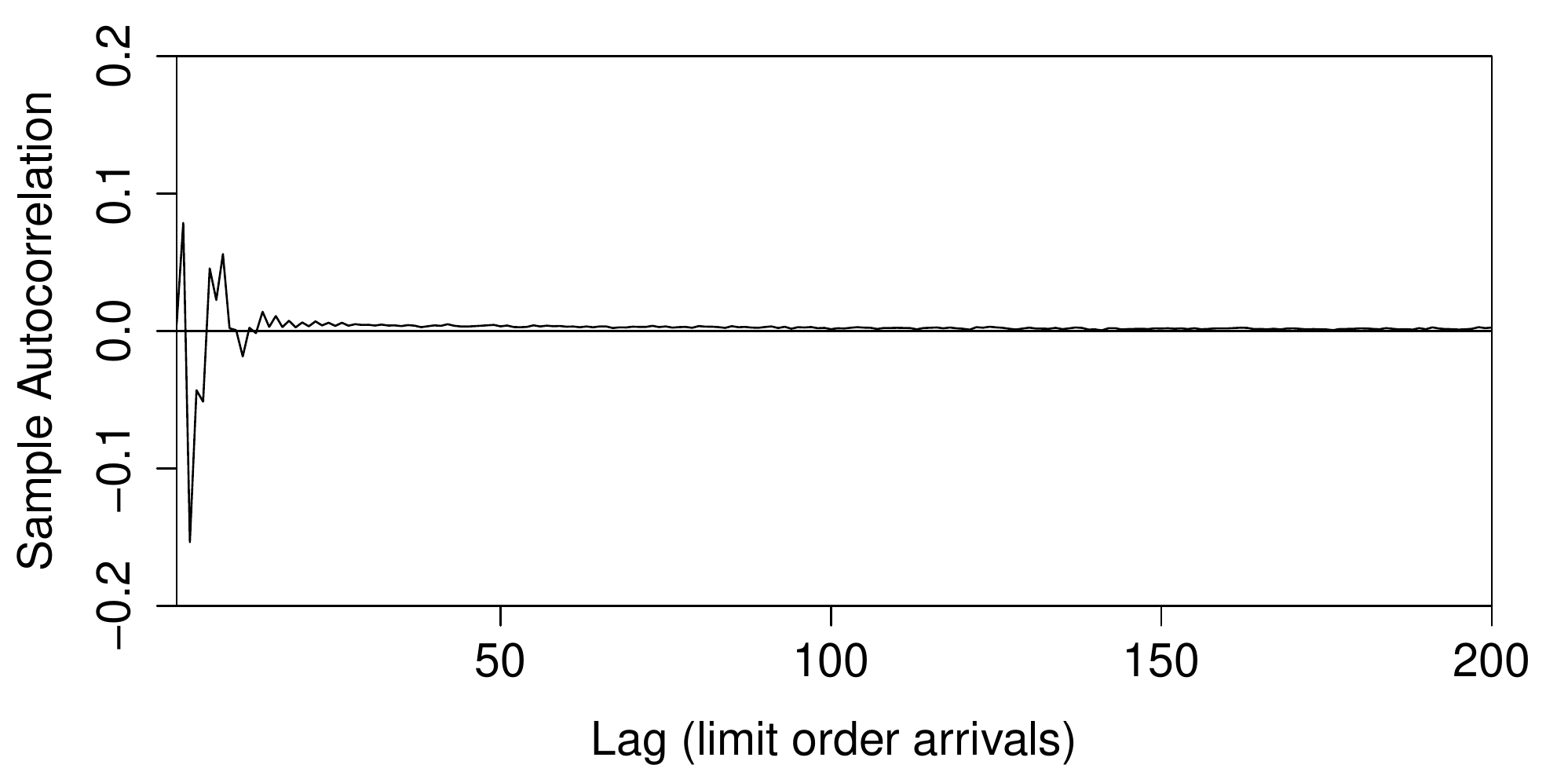}
\includegraphics[width=0.7\textwidth]{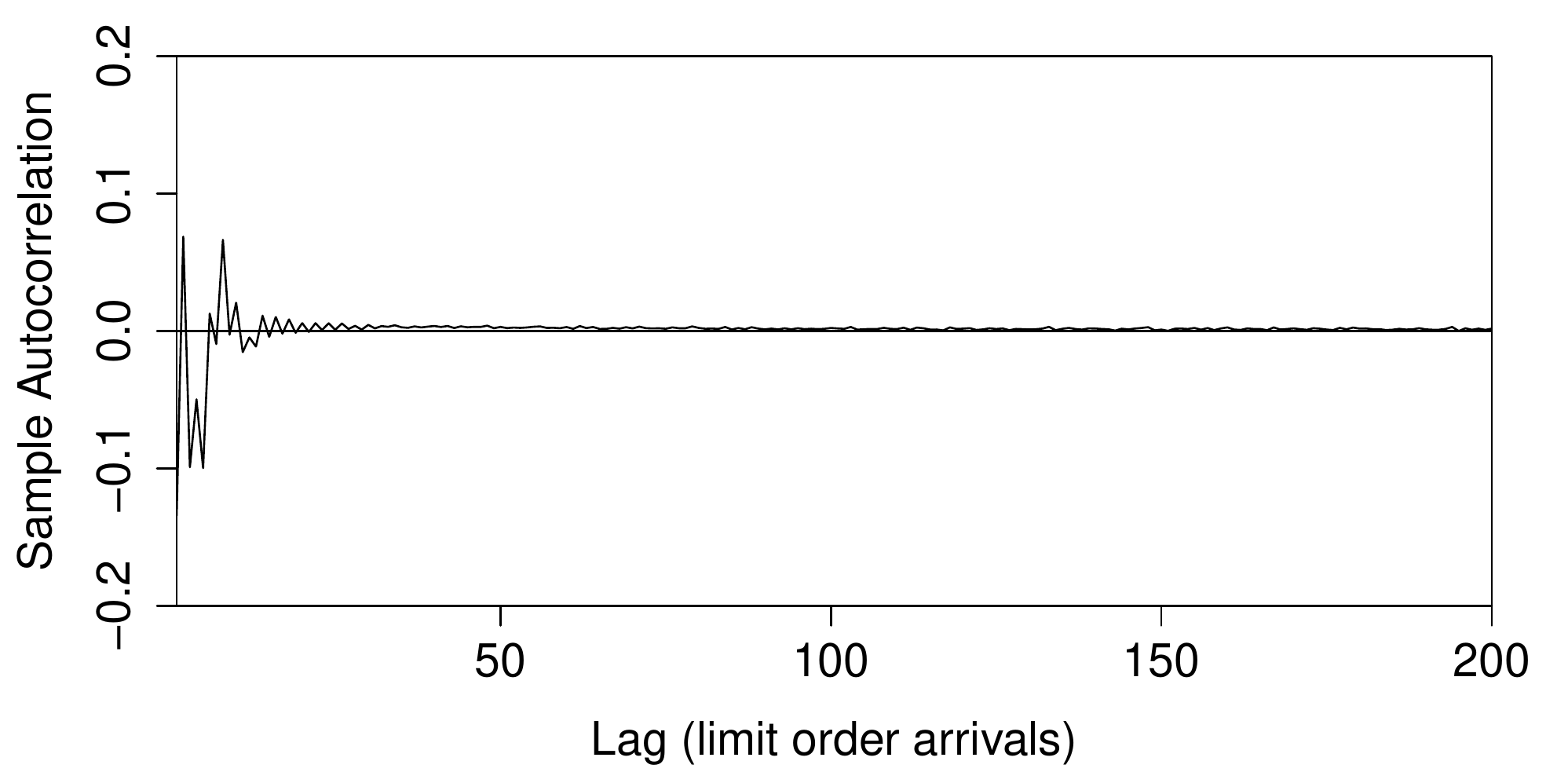}
\caption{Sample ACFs (see \ref{subsec:sampleacf}) for (top row) EUR/USD, (middle row) GBP/USD, and (bottom row) EUR/GBP intra-day arrival-sign series, which we construct using Equation (\ref{eq:loseries}). Each plot shows the sample ACF for 4 May 2010. The corresponding results for all other days in our sample and for the intra-day departure-sign series are qualitatively similar.}
\label{fig:ACFEU}
\end{figure}

In Figure \ref{fig:ACFEU}, we plot the sample ACFs (see \ref{subsec:sampleacf}) for each of the three currency pairs' intra-day arrival-sign series on 4 May 2010. The results for all other intra-day series on each day in our sample are qualitatively similar. Up to lags of about 25 events, the sample ACFs fluctuate between positive and negative values, which indicates that the order-flow series contain short-range negative autocorrelations. Although these autocorrelations have a magnitude below about $0.1$ and are therefore relatively weak, this effect is present on each day in our sample, and we therefore deem it to be a robust statistical property of the data. We return to our discussion of these negative autocorrelations in Section \ref{sec:lmdiscussion}.

\begin{figure}[!htbp]
\centering
\includegraphics[width=0.7\textwidth]{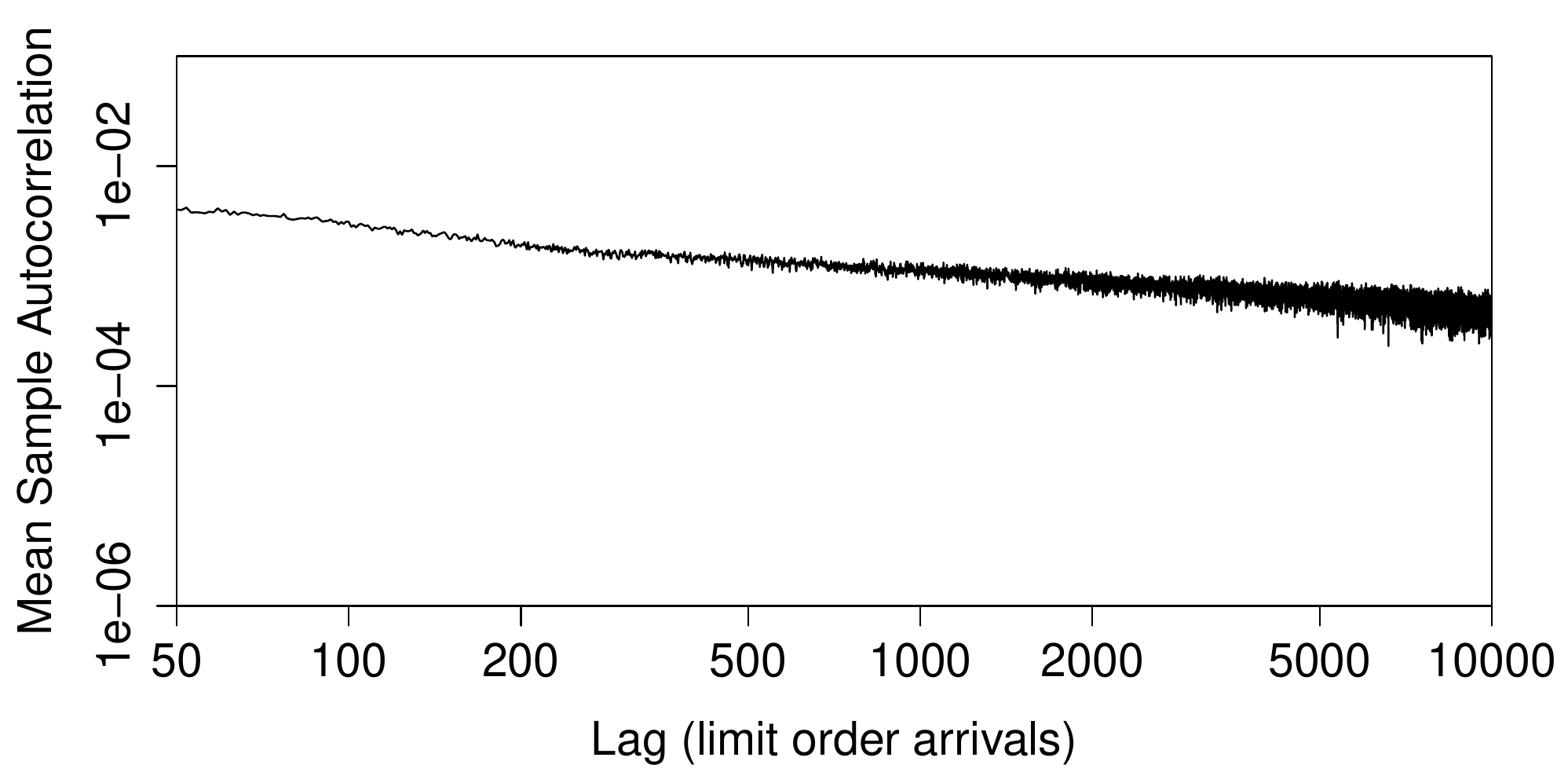}
\includegraphics[width=0.7\textwidth]{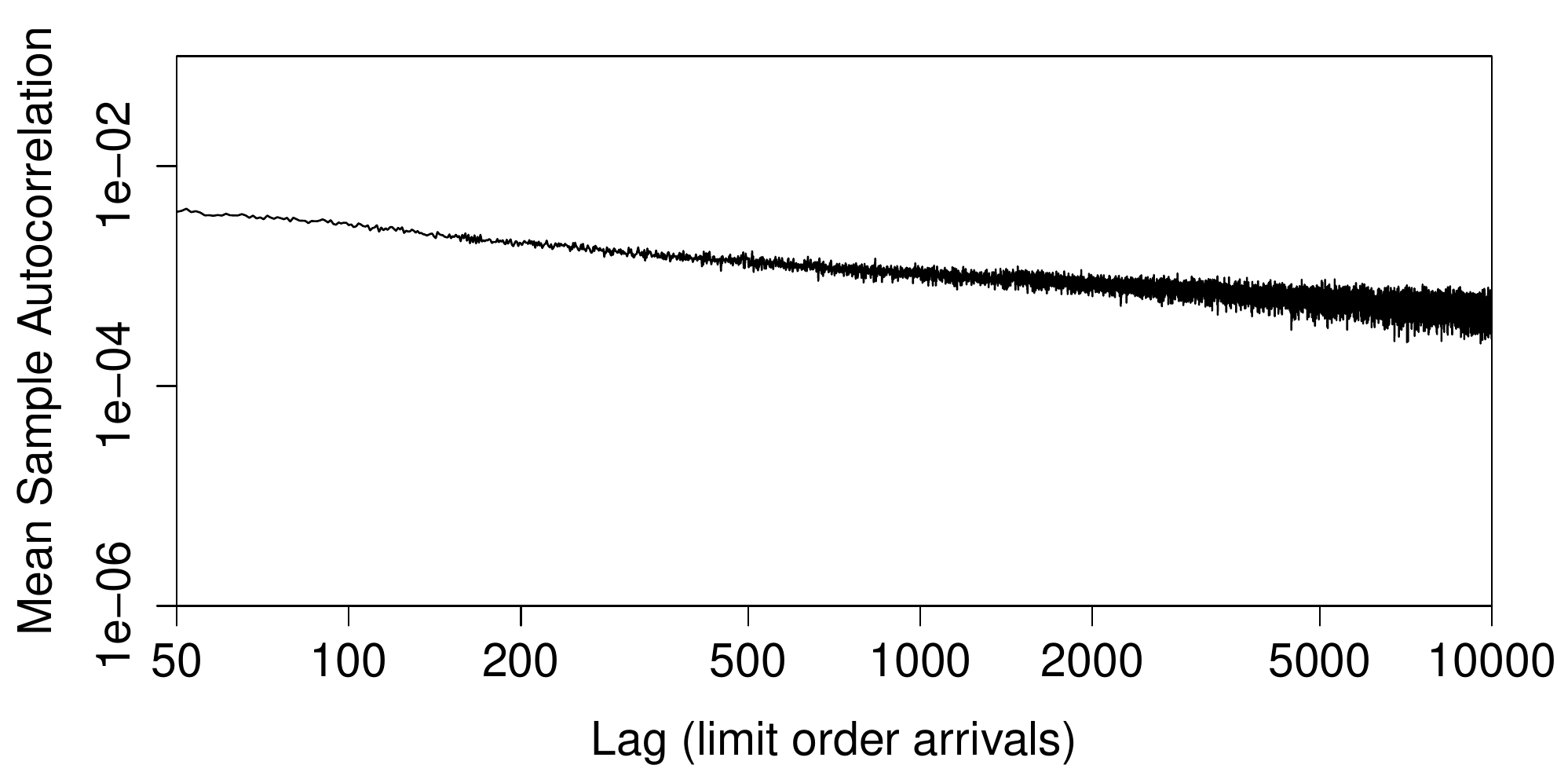}
\includegraphics[width=0.7\textwidth]{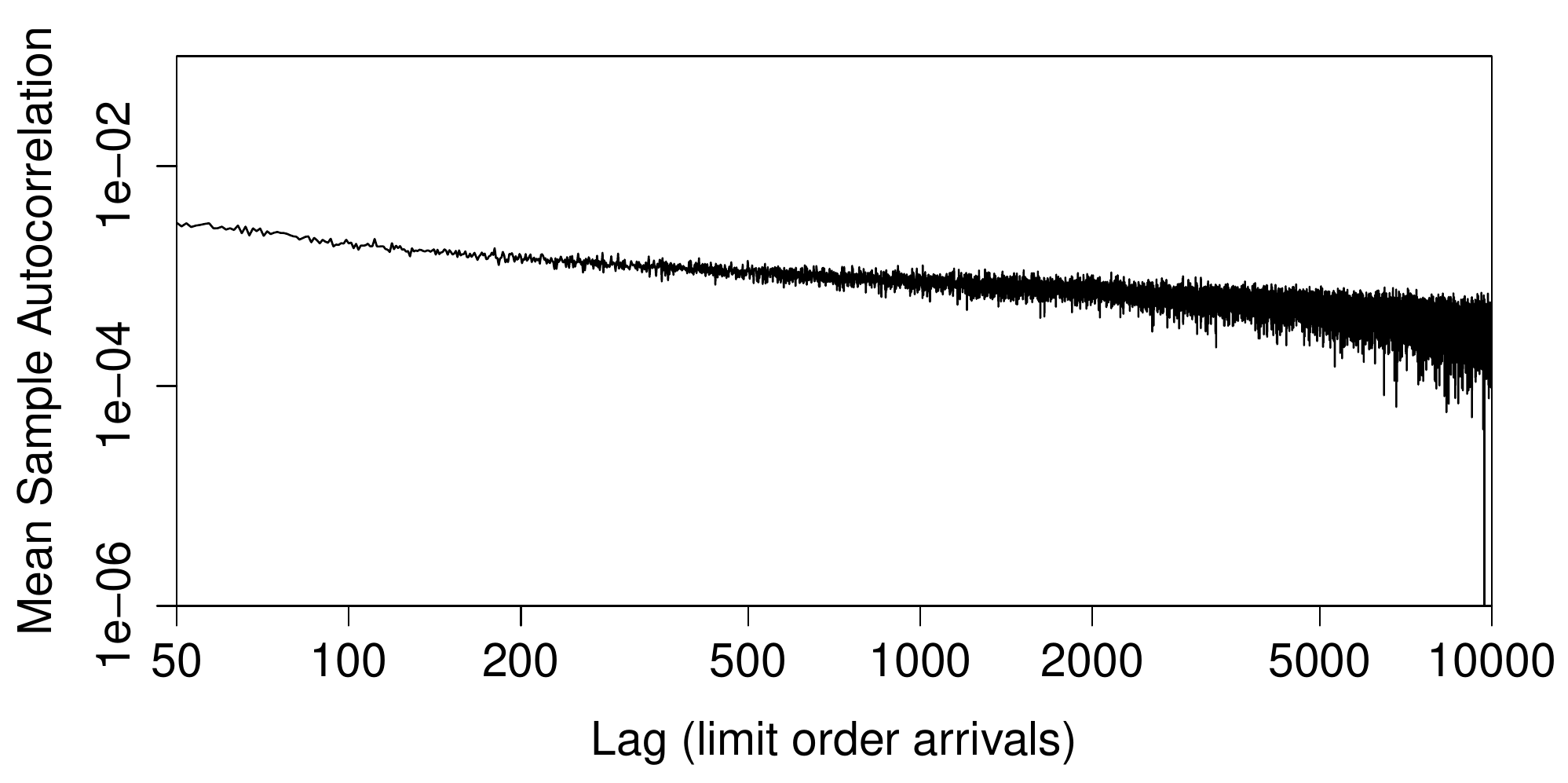}
\caption{Doubly logarithmic plots of mean sample ACFs (see \ref{subsec:sampleacf}) for (top row) EUR/USD, (middle row) GBP/USD, and (bottom row) EUR/GBP intra-day arrival-sign series. We obtain each plot by averaging the daily sample ACFs across all 30 days in our sample. We omit lags smaller than $50$ events because some values of the mean sample ACFs are negative in this range. The results for the intra-day departure-sign series are qualitatively similar.}
\label{fig:ACFEUloglog}
\end{figure}

In Figure \ref{fig:ACFEUloglog}, we plot the intra-day sample ACFs in doubly logarithmic coordinates. To help reduce the noise at higher lags,\footnote{The statistical errors associated with estimating the sample ACF are approximately constant at all lags, but the signal strength is smaller at larger lags because there are fewer independent data points \cite{Bouchaud:2009digest}.} we plot the mean sample ACFs, which we obtain by averaging the daily sample ACFs across all 30 days in our sample. After the short-term negative autocorrelations subside (which occurs before lag 50, the lower bound in our plots), the sample ACFs remain positive for lags of several thousands of events. This suggests that there are long-range, positive autocorrelations in the series.

\begin{figure}[!htbp]
\centering
\includegraphics[width=0.7\textwidth]{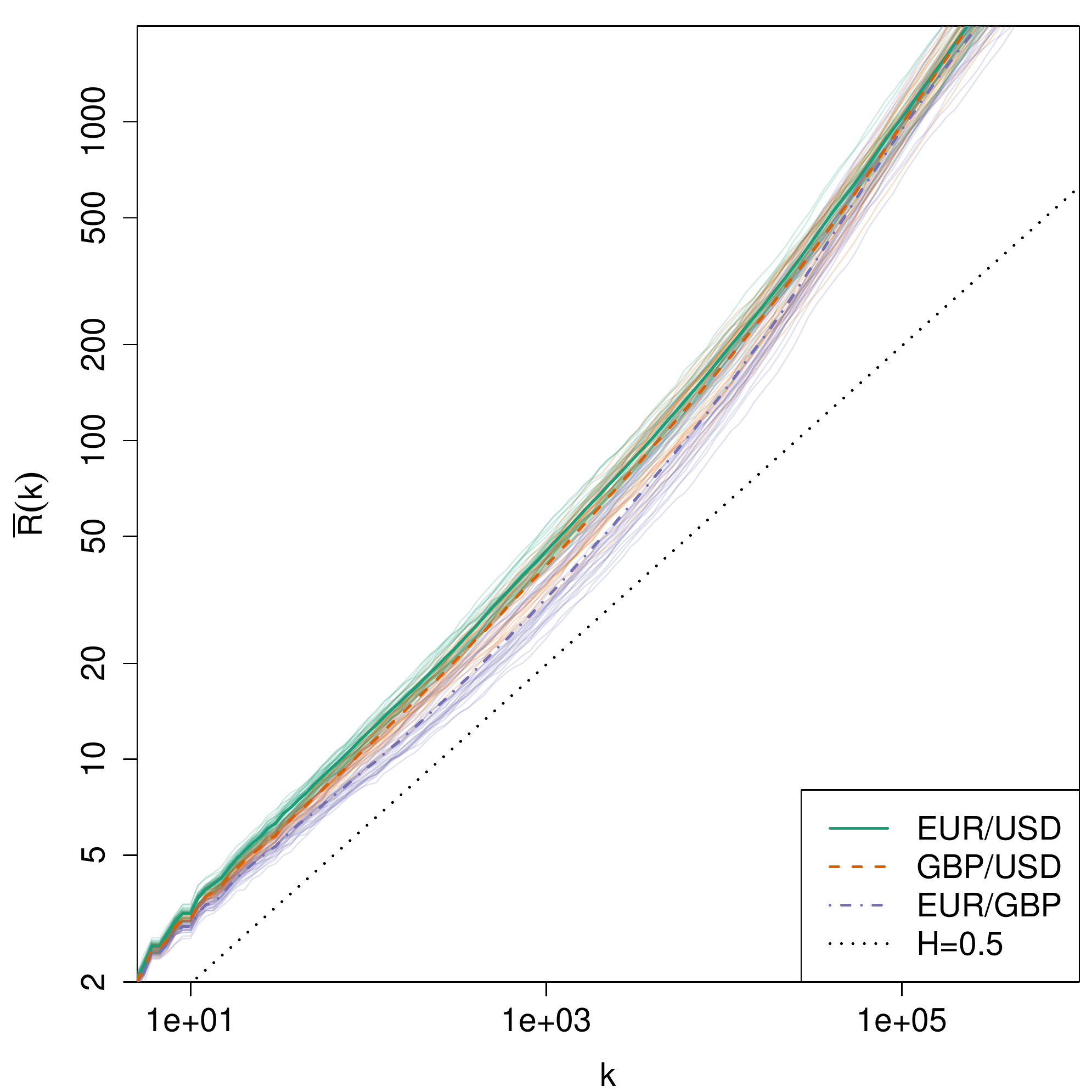}
\caption{Rescaled-range plots (see \ref{subsec:lmpox}) for (solid green curves) EUR/USD, (dashed orange curves) GBP/USD, and (dotted--dashed purple curves) EUR/GBP arrival-sign series. The pale curves indicate the rescaled-range statistics $\overline{R}(k)$ (see Equation (\ref{eq:poxmean})) for a single intra-day series, and the darker curves indicate the mean across all 30 intra-day series. The dotted black line has a slope of $0.5$. In these plots, we divide each intra-day series into $B=100$ blocks; we also produced similar plots for several different values of $B\in\left[10,1000\right]$ and obtained similar results. The results for the intra-day departure-sign series are qualitatively similar.}
\label{fig:POXEU}
\end{figure}

In Figure \ref{fig:POXEU}, we show rescaled-range plots (see \ref{subsec:lmpox}) for each of the three currency pairs' intra-day arrival-sign series. For each of the three currency pairs, the slope of the rescaled-range plot for each intra-day order-flow series is close to $0.5$ for values of $k$ below about $10000$. For larger values of $k$, the slope of each rescaled-range plot is above $0.5$. This suggests that the intra-day order-flow sign series are long-memory processes.

\begin{figure}[!htbp]
\centering
\includegraphics[width=0.8\textwidth]{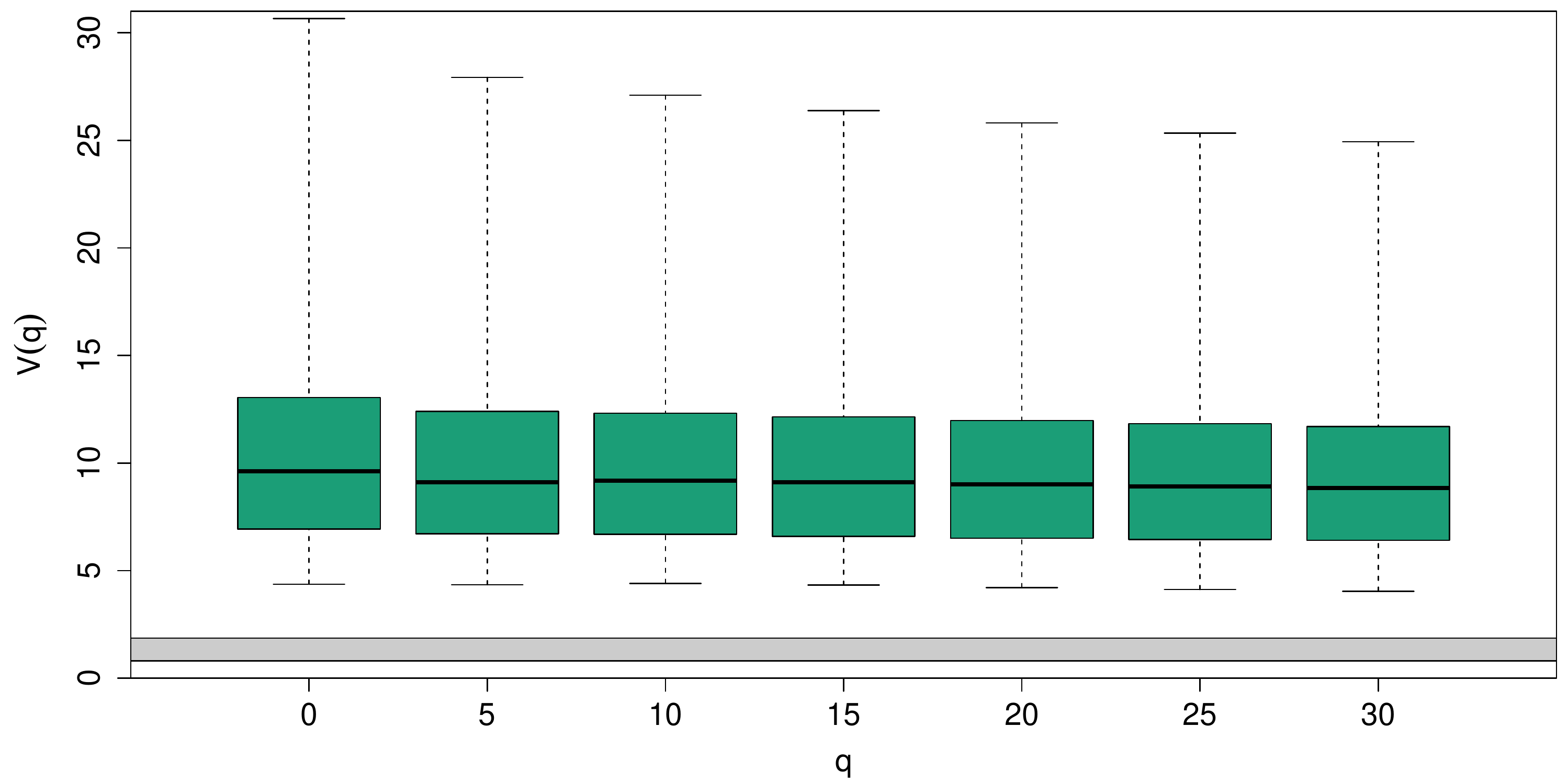}
\includegraphics[width=0.8\textwidth]{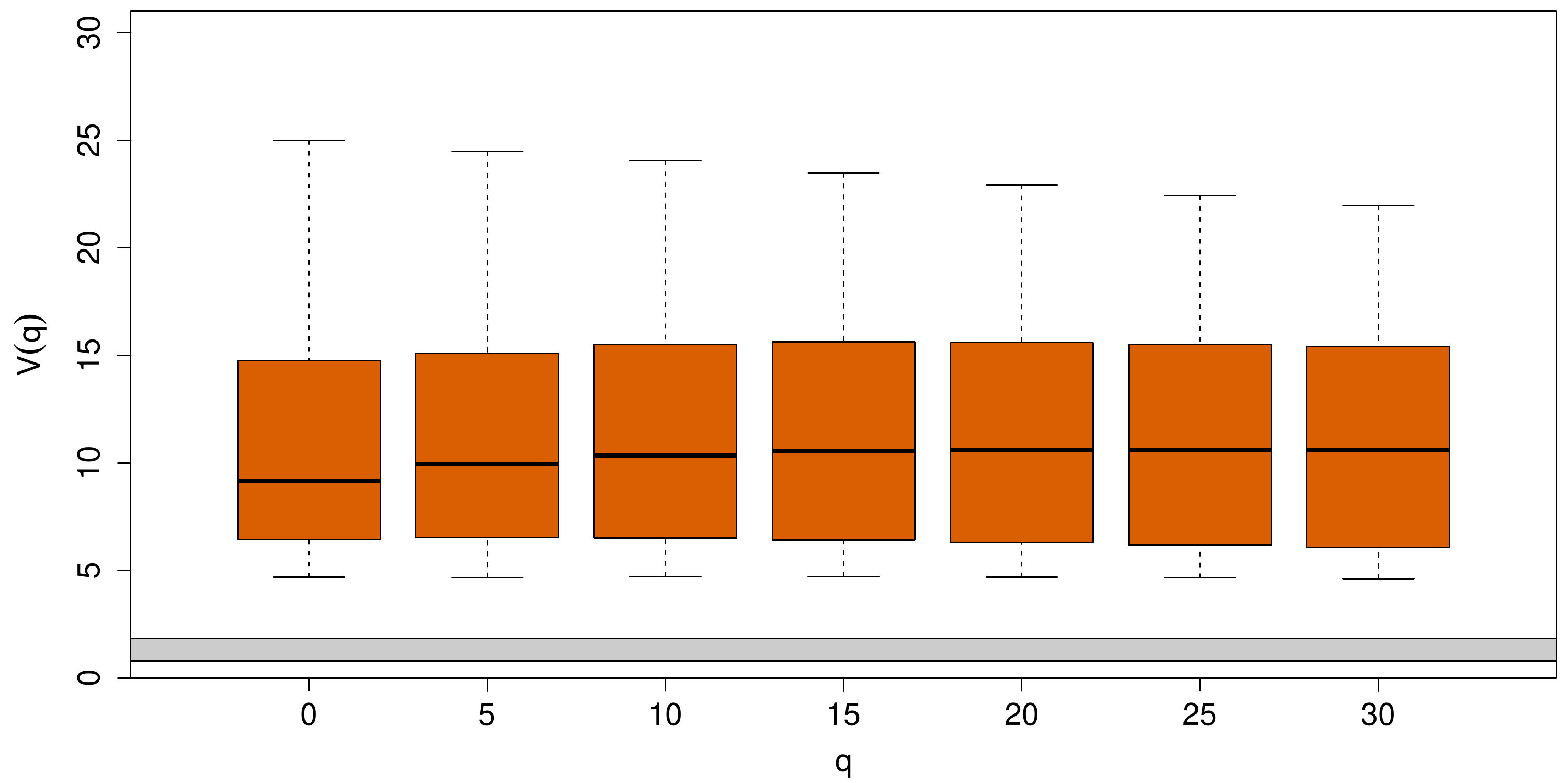}
\includegraphics[width=0.8\textwidth]{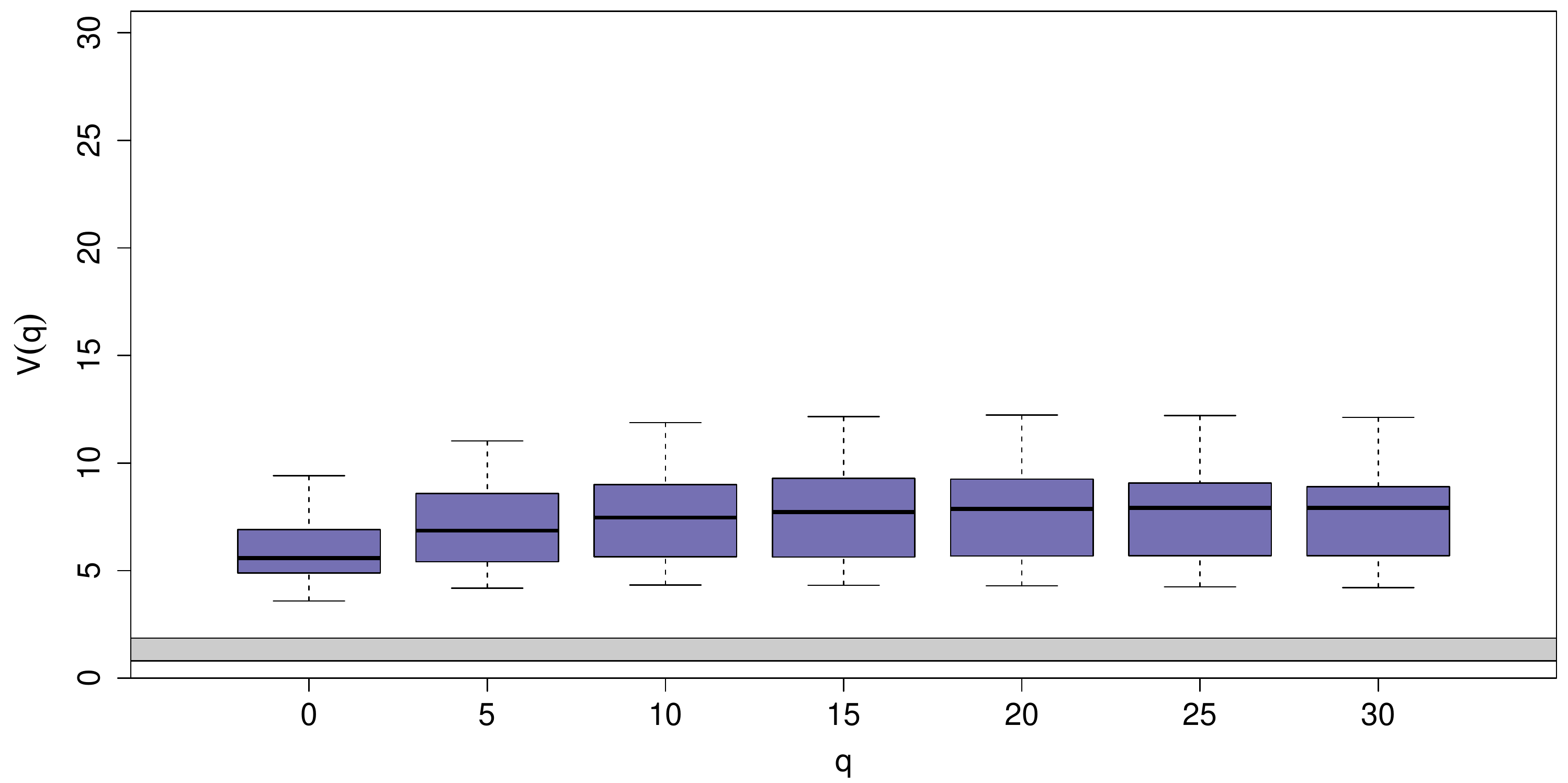}
\caption{Box plots of Lo's modified rescaled-range test statistic $V(q)$ (see \ref{subsec:lmlo}) for the given choices of bandwidth parameter $q$ for the (top) EUR/USD, (middle) GBP/USD, and (bottom) EUR/GBP intra-day arrival-sign series. For each choice of $q$, the boxes indicate the lower quartile, median, and upper quartile of $V(q)$ and the whiskers indicate the minimum and maximum of $V(q)$, across all 30 intra-day series. The light grey shading indicates the critical region for Lo's modified rescaled-range test at the $5\%$ significance level. The results for the intra-day departure-sign series are qualitatively similar.}
\label{fig:LOTESTRLC}
\end{figure}

To test the hypothesis of long memory more formally, we perform Lo's modified rescaled-range test (see \ref{subsec:lmlo}) on the intra-day arrival series (see Figure \ref{fig:LOTESTRLC}). For each day in our sample, Lo's modified rescaled-range test causes us to reject the null hypothesis of short memory at the $5\%$ significance level at all bandwidth choices that we study. Similarly, when using Andrews' plug-in estimator (see Equation (\ref{eq:lmandrewsqhat})) to estimate a suitable choice of bandwidth parameter from the data, Lo's test rejects the null hypothesis of short memory for each of the three currency pairs and on all 30 days in our sample. Therefore, Lo's test provides strong evidence that the intra-day order-sign series are long-memory processes.

Given the strong results of Lo's test, we now turn to assessing the strength of the long memory in intra-day order flow. To do so, we use two different methods to estimate the Hurst exponent $H$ (see \ref{subsec:hurst}): DFA (see \ref{subsec:lmdfa}) and log-periodogram regression (see \ref{subsec:lmlp}). Due to the negative short-range autocorrelations that we observe in the sample ACFs (see Figure \ref{fig:ACFEU}), it is necessary to identify sensible choices of input parameters --- namely, the minimal window length $m_{\min}$ of a DFA and the number $c$ of Fourier frequencies in a log-periodogram regression --- when performing these estimation techniques.

\begin{figure}[!htbp]
\centering
\includegraphics[width=0.7\textwidth]{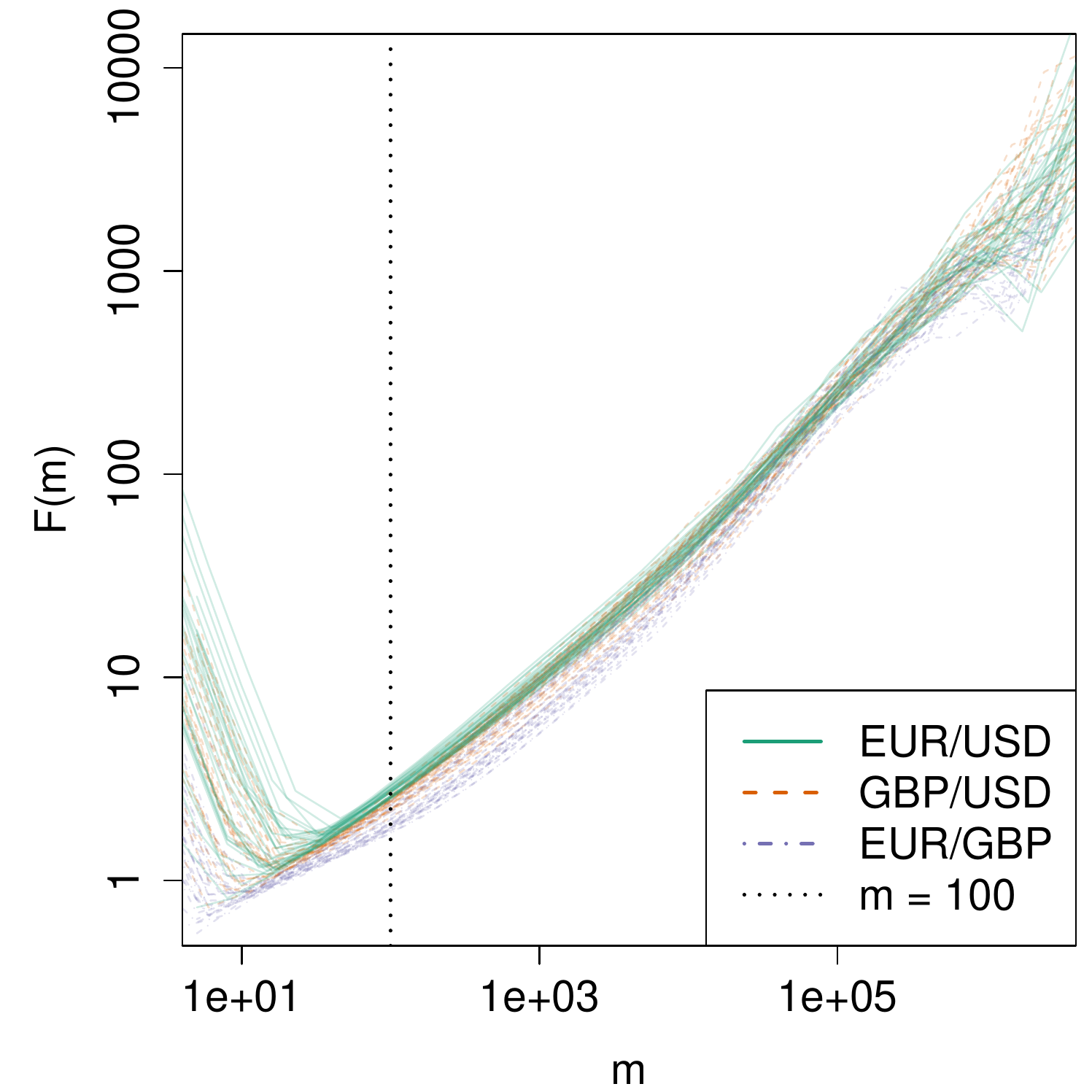}
\caption{Length-$m$ mean detrended standard deviation $F(m)$ (see \ref{subsec:lmdfa}) for the (solid green curves) EUR/USD, (dashed orange curves) GBP/USD, and (dotted--dashed purple curves) EUR/GBP arrival-sign series. Each curve corresponds to a single intra-day series. The dotted black line indicates $m=100$. The results for the intra-day departure-sign series are qualitatively similar.}
\label{fig:DFAlines}
\end{figure}

To identify a suitable choice of $m_{\min}$ for our DFA estimates of $H$, we first plot the length-$m$ mean detrended standard deviation $F(m)$ for several choices of $m$ (see Figure \ref{fig:DFAlines}). For window lengths $m$ that are smaller than about 25, the negative autocorrelations dominate the mean detrended standard deviations $F(m)$. These values of $m$ are therefore unsuitable for calculating a DFA estimate of $H$. For values of $m$ larger than about 100, the log-log plots of $F(m)$ follow an approximately straight line. We therefore perform our DFA estimates of $H$ using $m_{\min}=100$.

\begin{figure}[!htbp]
\centering
\includegraphics[width=0.7\textwidth]{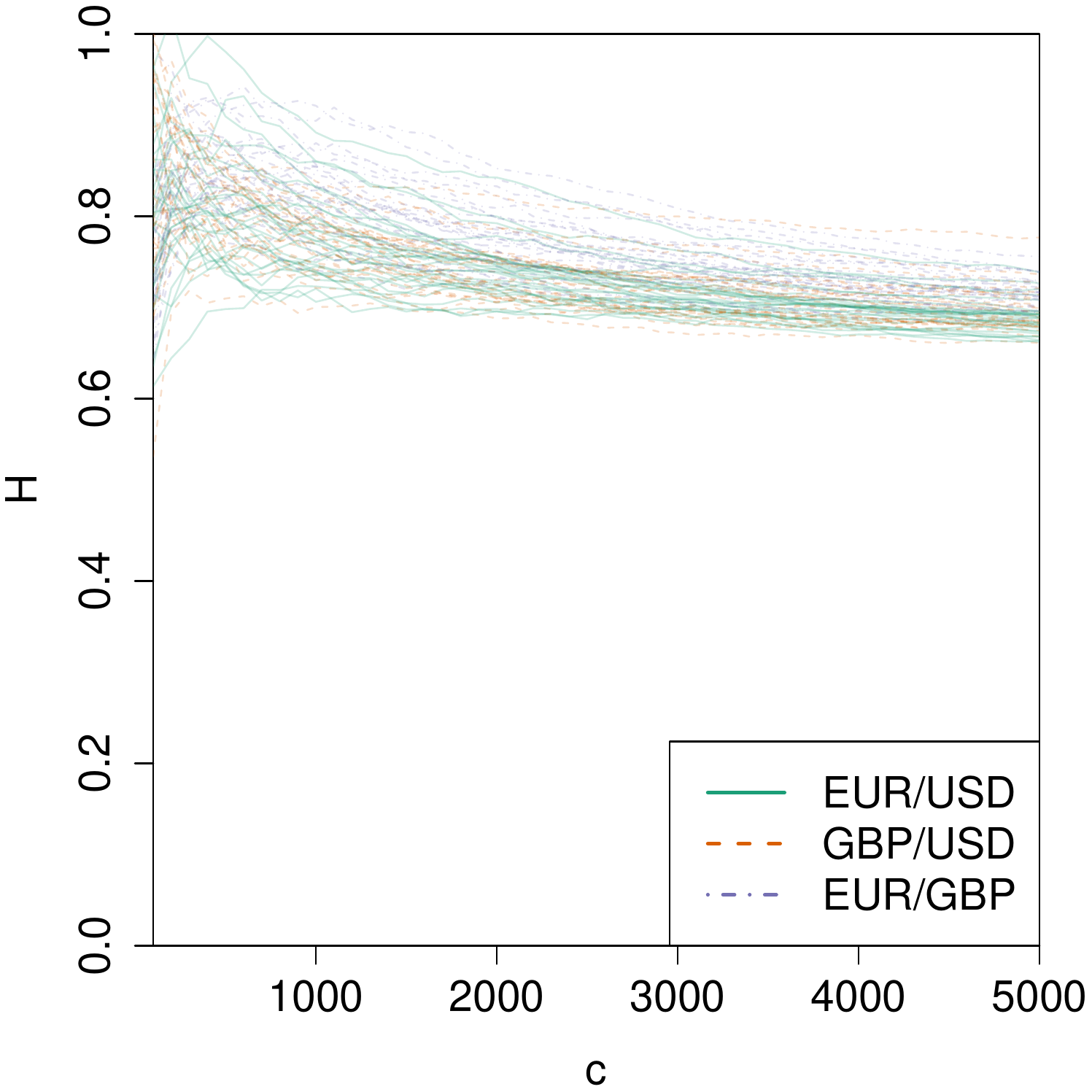}
\caption{Log-periodogram regression (see \ref{subsec:lmlp}) estimates of $H$ for given number $c$ of Fourier frequencies for the (solid green curves) EUR/USD, (dashed orange curves) GBP/USD, and (dotted--dashed purple curves) EUR/GBP arrival-sign series. Each curve corresponds to a single intra-day series. The results for the intra-day departure-sign series are qualitatively similar.}
\label{fig:GPHlines}
\end{figure}

To identify a suitable choice of $c$ for our log-periodogram regression estimates of $H$, we plot the log-periodogram regression estimates of $H$ for several different values of $c$ (see Figure \ref{fig:GPHlines}). In all cases, the estimates of $H$ tend to decrease as $c$ increases, and there is no clear plateau over which the estimates of $H$ are stable. In the absence of an obvious choice for $c$, we use the popular rule-of-thumb \cite{Geweke:1983estimation} $c=\sqrt{N}$, where $N$ is the length of the given series. We stress, however, that the plots in Figure \ref{fig:GPHlines} indicate that our log-periodogram regression estimates of $H$ depend heavily on this choice, so using a different choice for $c$ would produce quantitatively different results. For example, another popular rule-of-thumb \cite{Taqqu:1995estimators} is $c=0.1\times(N/2)$. Due to the extremely large size of the intra-day order-flow series, this choice produces estimates of $H \approx 0.5$, which we do not regard to be sensible given that the other statistical tests all suggest that the intra-day order-flow series exhibit long memory. The absence of a clear choice for $c$ highlights a weakness of log-periodogram regression for the present application.

\begin{figure}[!htbp]
\centering
\includegraphics[width=0.7\textwidth]{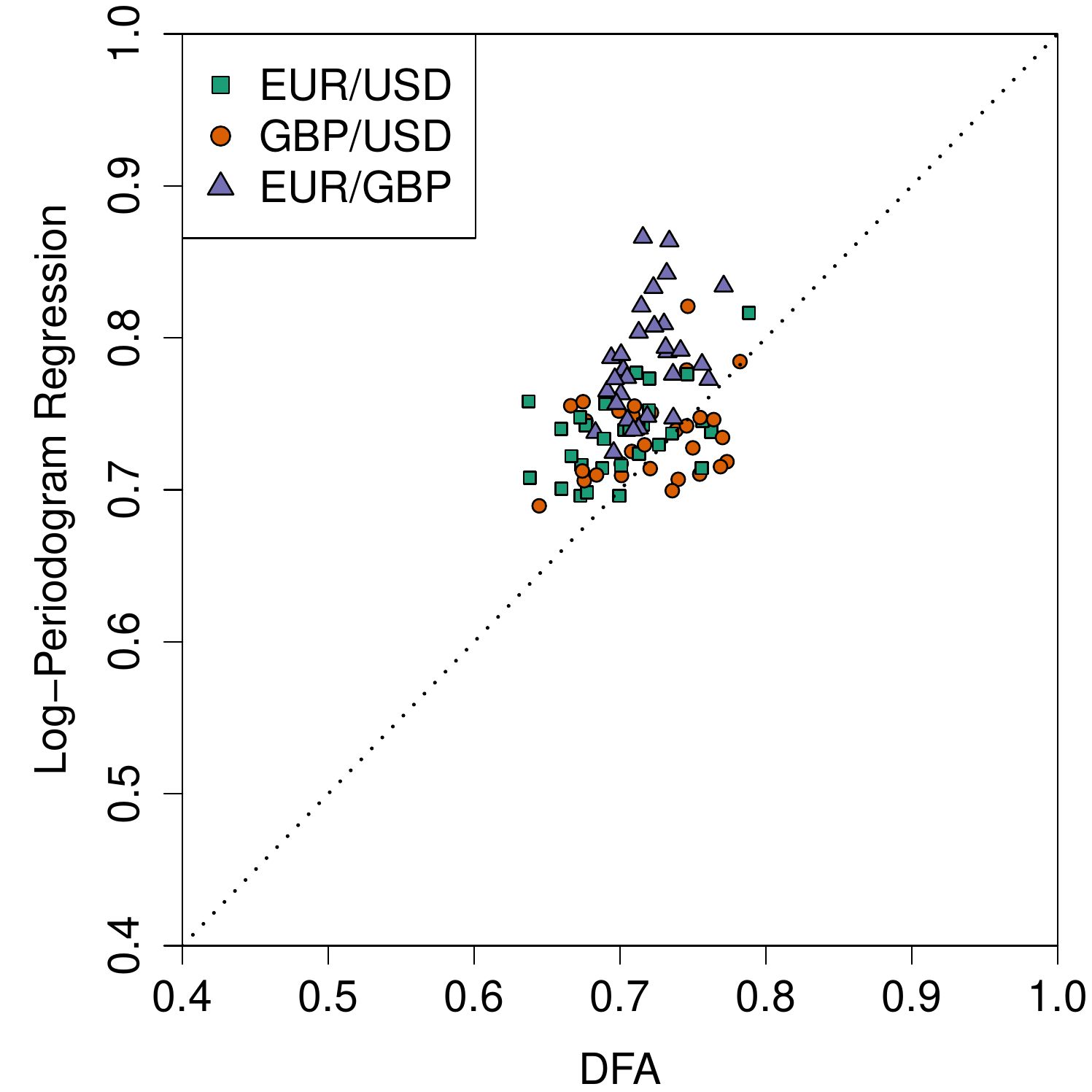}
\caption{DFA and log-periodogram regression estimates of the Hurst exponent $H$ for the intra-day (green squares) EUR/USD, (orange circles) GBP/USD, and (purple triangles) EUR/GBP arrival-sign series. Each point corresponds to the estimates for a single trading day. The dotted black line indicates the diagonal. The results for the intra-day departure-sign series are qualitatively similar.}
\label{fig:HurstCORR}
\end{figure}

In Figure \ref{fig:HurstCORR}, we plot the DFA and log-periodogram regression estimates of $H$ (using our choices of $m_{\min}=100$ and $c=\sqrt{N}$) for each intra-day arrival-sign series. For each of the three currency pairs, the DFA estimates of $H$ cluster in the range of about $0.6$ to about $0.8$. The log-periodogram regression estimates of $H$ tend to be slightly larger (they cluster in the range of about $0.65$ to about $0.85$). However, the latter results depend heavily on the choice of $c$ (see Figure \ref{fig:GPHlines}), so we deem the DFA estimates to be more useful.

In Table \ref{tab:Hmeans}, we list the means and standard deviations of our DFA and log-periodogram regression estimates of $H$ across all 30 days in our sample. As we noted above, our results for arrival-sign series and departure-sign series are very similar. In all cases, the mean estimates of $H$ are many standard deviations larger than $0.5$, which strongly supports the hypothesis that these series exhibit long memory. Based on our results in Figures \ref{fig:DFAlines}, \ref{fig:GPHlines}, and \ref{fig:HurstCORR} and Table \ref{tab:Hmeans}, we regard $H \approx 0.7$ to be a good estimate for the Hurst exponent of the arrival-sign series and departure-sign series for each of the three currency pairs.

\begin{table}[!htbp]
\begin{center}
\begin{tabular}{|c|c|c|c|}
\hline
& & DFA & Log-Periodogram Regression \\
\hline
\multirow{2}{*}{EUR/USD} & Arrivals & $0.70$ $(0.04)$ & $0.74$ $(0.03)$ \\ 
& Departures & $0.70$ $(0.04)$ & $0.74$ $(0.03)$ \\
\hline
\multirow{2}{*}{GBP/USD} & Arrivals & $0.72$ $(0.04)$ & $0.74$ $(0.03)$ \\ 
& Departures & $0.72$ $(0.03)$ & $0.74$ $(0.03)$ \\ 
\hline
\multirow{2}{*}{EUR/GBP} & Arrivals & $0.72$ $(0.02)$ & $0.79$ $(0.04)$ \\ 
& Departures & $0.71$ $(0.02)$ & $0.78$ $(0.04)$ \\ 
\hline
\end{tabular}
\caption{DFA and log-periodogram regression estimates of the Hurst exponent $H$ for the EUR/USD, GBP/USD, and EUR/GBP intra-day arrival-sign and departure-sign series. Each entry indicates the mean of the estimates across all intra-day series. The numbers in parentheses indicate 1 standard deviation of the estimates across all intra-day series.}
\label{tab:Hmeans}
\end{center}
\end{table}

\subsection{Results for Cross-Day Series}\label{subsec:lmcrossday}

To assess how aggregating data from different trading days impacts our results, we now repeat all of our calculations using order-flow series that span daily boundaries. Specifically, for a pair of consecutive trading days $D_i$ and $D_{i+1}$, we construct the \emph{cross-day arrival-sign series} by concatenating the second half\footnote{If an intra-day series has odd length, we round down to the previous integer.} of the intra-day arrival-sign series from day $D_i$ and the first half of the corresponding intra-day arrival-sign series from day $D_{i+1}$. We construct the \emph{cross-day departure-sign series} similarly using the corresponding departure-sign series.

\begin{figure}[!htbp]
\centering
\includegraphics[width=0.7\textwidth]{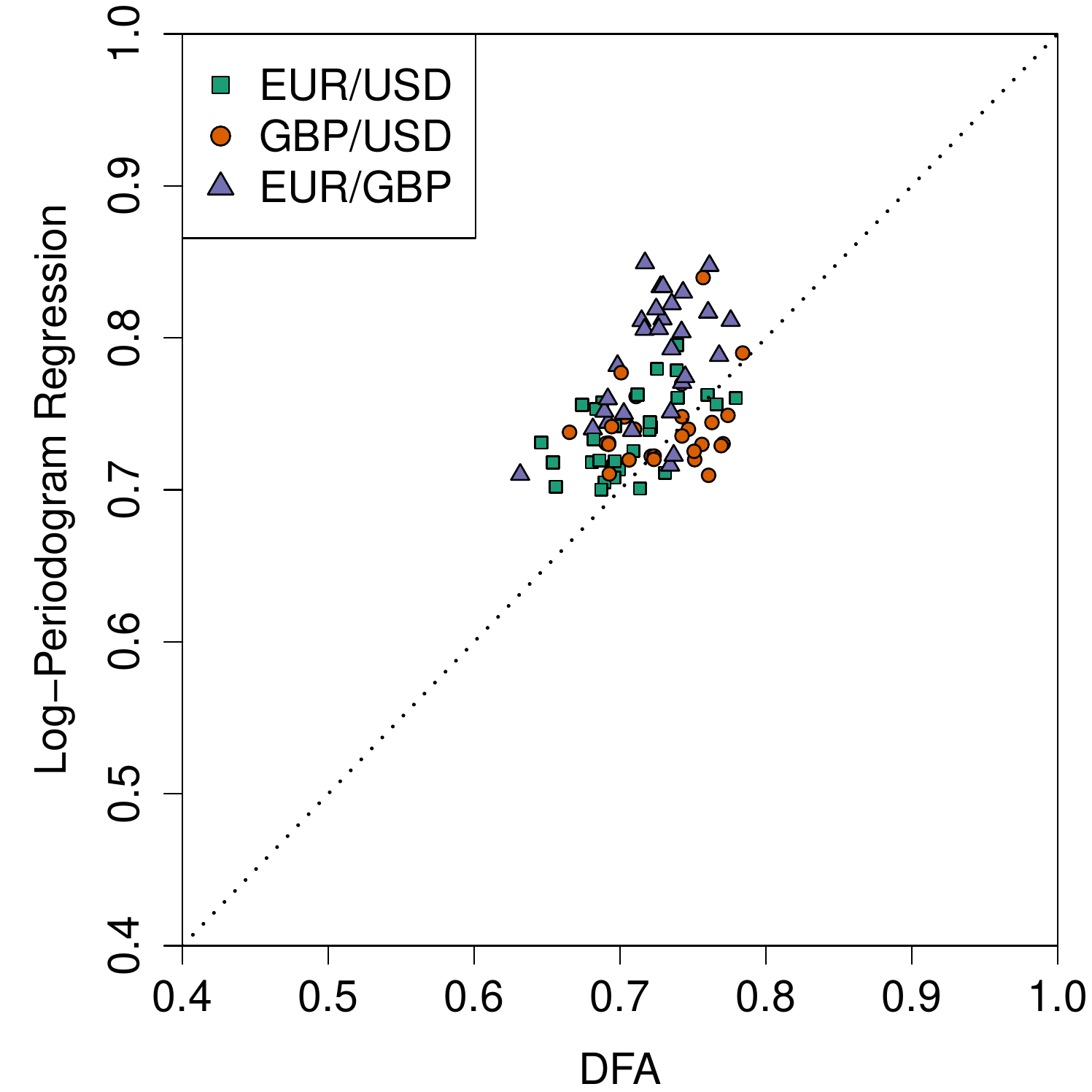}
\caption{DFA and log-periodogram regression estimates of the Hurst exponent $H$ for the cross-day (green squares) EUR/USD, (orange circles) GBP/USD, and (purple triangles) EUR/GBP arrival-sign series. Each point corresponds to the estimates for a consecutive pair of trading days. The dotted black line indicates the diagonal. The results for the cross-day departure-sign series are qualitatively similar.}
\label{fig:HCORR_CD}
\end{figure}

In all cases, we find that the sample ACFs, rescaled-range plots, and results from Lo's modified rescaled-range test are qualitatively similar to those for the intra-day series (see Figures \ref{fig:ACFEU}, \ref{fig:ACFEUloglog}, \ref{fig:POXEU}, and \ref{fig:LOTESTRLC}). This provides strong evidence that the cross-day order-flow series exhibit long memory. To quantify the strength of this long memory, we calculate DFA and log-periodogram regression estimates of the Hurst exponent $H$ using the same choices of input parameters ($m_{\min}=100$ and $c = \sqrt{N}$) as we used for our corresponding estimates of $H$ for the intra-day series. We plot our results in Figure \ref{fig:HCORR_CD} and list the means and standard deviations of our estimates across all 29 cross-day periods in Table \ref{tab:HmeansCD}.

\begin{table}[!htbp]
\begin{center}
\begin{tabular}{|c|c|c|c|}
\hline
& & DFA & Log-Periodogram Regression \\
\hline
\multirow{2}{*}{EUR/USD} & Arrivals & $0.71$ $(0.03)$ & $0.74$ $(0.03)$ \\ 
& Departures & $0.70$ $(0.03)$ & $0.74$ $(0.03)$ \\
\hline
\multirow{2}{*}{GBP/USD} & Arrivals & $0.73$ $(0.03)$ & $0.74$ $(0.03)$ \\ 
& Departures & $0.73$ $(0.03)$ & $0.74$ $(0.03)$ \\ 
\hline
\multirow{2}{*}{EUR/GBP} & Arrivals & $0.72$ $(0.03)$ & $0.79$ $(0.04)$ \\ 
& Departures & $0.72$ $(0.03)$ & $0.78$ $(0.04)$ \\ 
\hline
\end{tabular}
\caption{DFA and log-periodogram regression estimates of the Hurst exponent $H$ for the EUR/USD, GBP/USD, and EUR/GBP cross-day arrival-sign and departure-sign series. Each entry indicates the mean of the estimates across all cross-day series. The numbers in parentheses indicate 1 standard deviation of the estimates across all cross-day series.}
\label{tab:HmeansCD}
\end{center}
\end{table}

Our estimates of $H$ for the cross-day series are very similar to the corresponding estimates for the intra-day series (see Figure \ref{fig:HCORR_CD} and Table \ref{tab:Hmeans}). As with the intra-day series, the DFA estimates of $H$ cluster in the range of about $0.6$ to about $0.8$ and the log-periodogram regression estimates of $H$ cluster in the range of about $0.65$ to about $0.85$. We therefore regard $H \approx 0.7$ to be a good estimate for the Hurst exponent of the cross-day order-sign series for each of the three currency pairs.

\subsection{Long Memory versus Nonstationarity}

In this section, we address the conjecture by Axioglou and Skouras \cite{Axioglou:2011markets} that the apparent long memory in the cross-day series is mostly an artifact caused by structural breaks at the boundaries between different trading days (see Section \ref{sec:litrev}).

To assess whether the estimator is able to identify the boundary between different trading days, we calculate the normalized cumulative-sum change-point estimate (see Equation (\ref{eq:berkesk})) for each cross-day order flow series . Because the lengths of the intra-day series vary with the number of arrivals and departures each day, the locations of the boundaries between different trading days vary across different cross-day series. Consequently, the boundary between trading days $D_i$ and $D_{i+1}$ does not necessarily lie at the mid point in the relevant cross-day series. We therefore introduce a normalization to enable comparisons between different cross-day series. Given a cross-day series of length $N$ with daily boundary $r^* \in \left\{2,3,\ldots,N-1\right\}$, and given a change-point estimator $\hat{r}^*$ for $r^*$, the \emph{normalized change-point estimator} is\begin{equation}\label{eq:rtilde}\tilde{r} = \left\{\begin{array}{ll}(\hat{r}^*-r^*)/r^*,& \text{ if }\hat{r}^*\leq r^*,\\(\hat{r}^*-r^*)/(r-r^*),& \text{ otherwise.}\end{array}\right.\end{equation}Observe that $\tilde{r}\in \left[-1,1\right]$, and $\tilde{r} =0$ if and only if $\hat{r}^*=r^*$.

\begin{figure}[!htbp]
\centering
\includegraphics[width=0.7\textwidth]{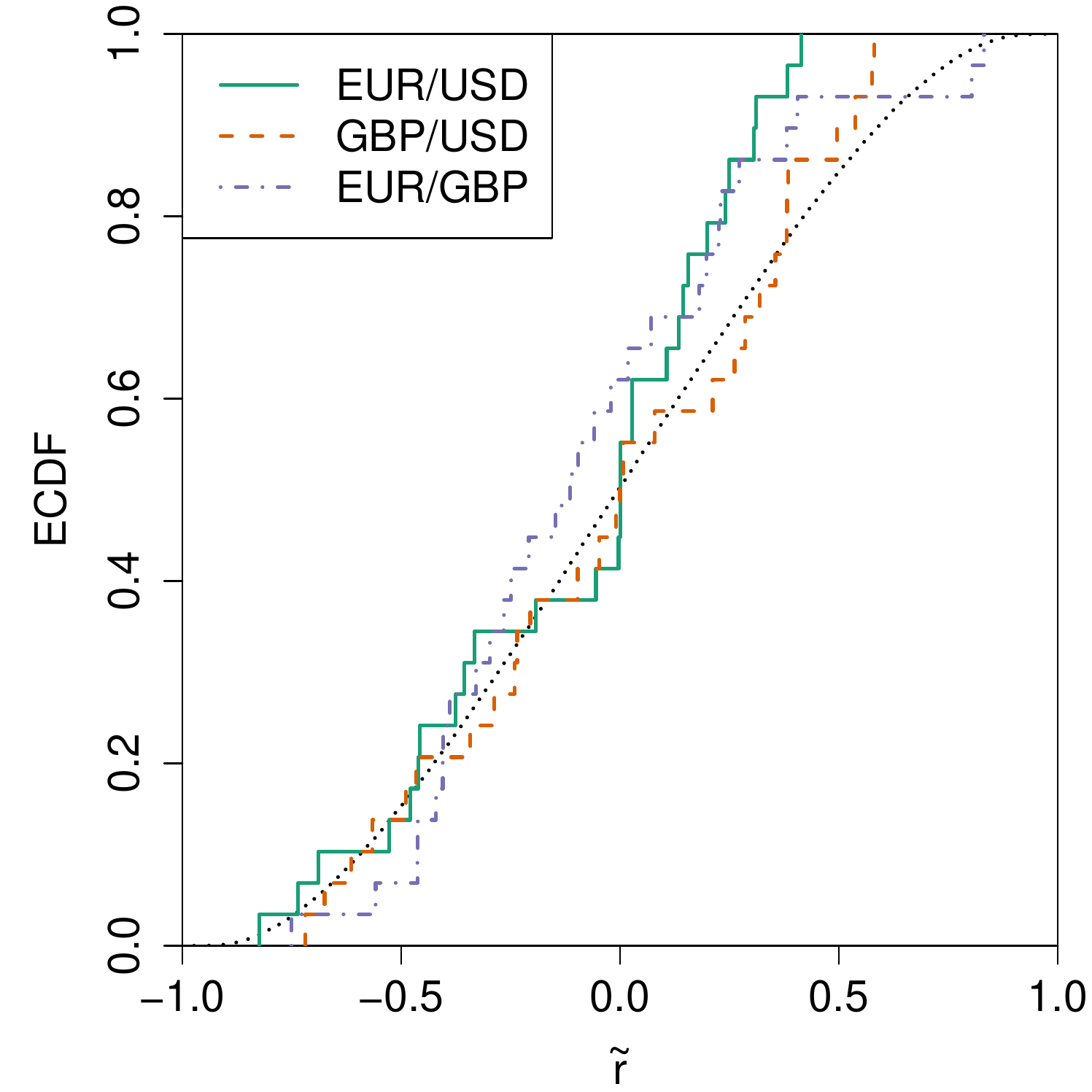}
\caption{ECDFs of the normalized cumulative-sum change-point estimator $\tilde{r}$ (see Equation (\ref{eq:rtilde}) for the (solid green curve) EUR/USD, (dashed orange curve) GBP/USD, and (dotted--dashed purple curve) EUR/GBP cross-day arrival-sign series. The results for the cross-day departure-sign series are qualitatively similar. The dotted black curve illustrates the estimator's null distribution for a second-order stationary series with no structural breaks, which we estimate by calculating the ECDF of the normalized cumulative-sum change-point estimates from $100000$ independent series that each consist of $1000000$ random variables drawn independently and at random from the standard normal distribution.}
\label{fig:CUSUMk}
\end{figure}

In Figure \ref{fig:CUSUMk}, we plot the empirical cumulative density functions (ECDFs) of $\tilde{r}$ for each of the three currency pairs' cross-day arrival-sign series. The results for the cross-day departure-sign series are qualitatively similar. For the EUR/GBP cross-day series, there is no visible jump in the ECDF near $\tilde{r}=0$. Therefore, the cumulative-sum change-point estimator performs very poorly at detecting the true locations of the daily boundaries in the cross-day series for this currency pair. For EUR/USD and GBP/USD, there is a small but discernible spike in the ECDFs near to $\tilde{r}=0$, which indicates that the estimator performs somewhat better for these currency pairs. However, the distribution of the estimator's output across different trading days is still very broad, which suggests that its performance is still rather weak. Together, these results suggest that the apparent long-memory properties of the cross-day series are not strongly influenced by nonstationarities at the daily boundaries.

\begin{figure}[!htbp]
\centering
\includegraphics[width=0.49\textwidth]{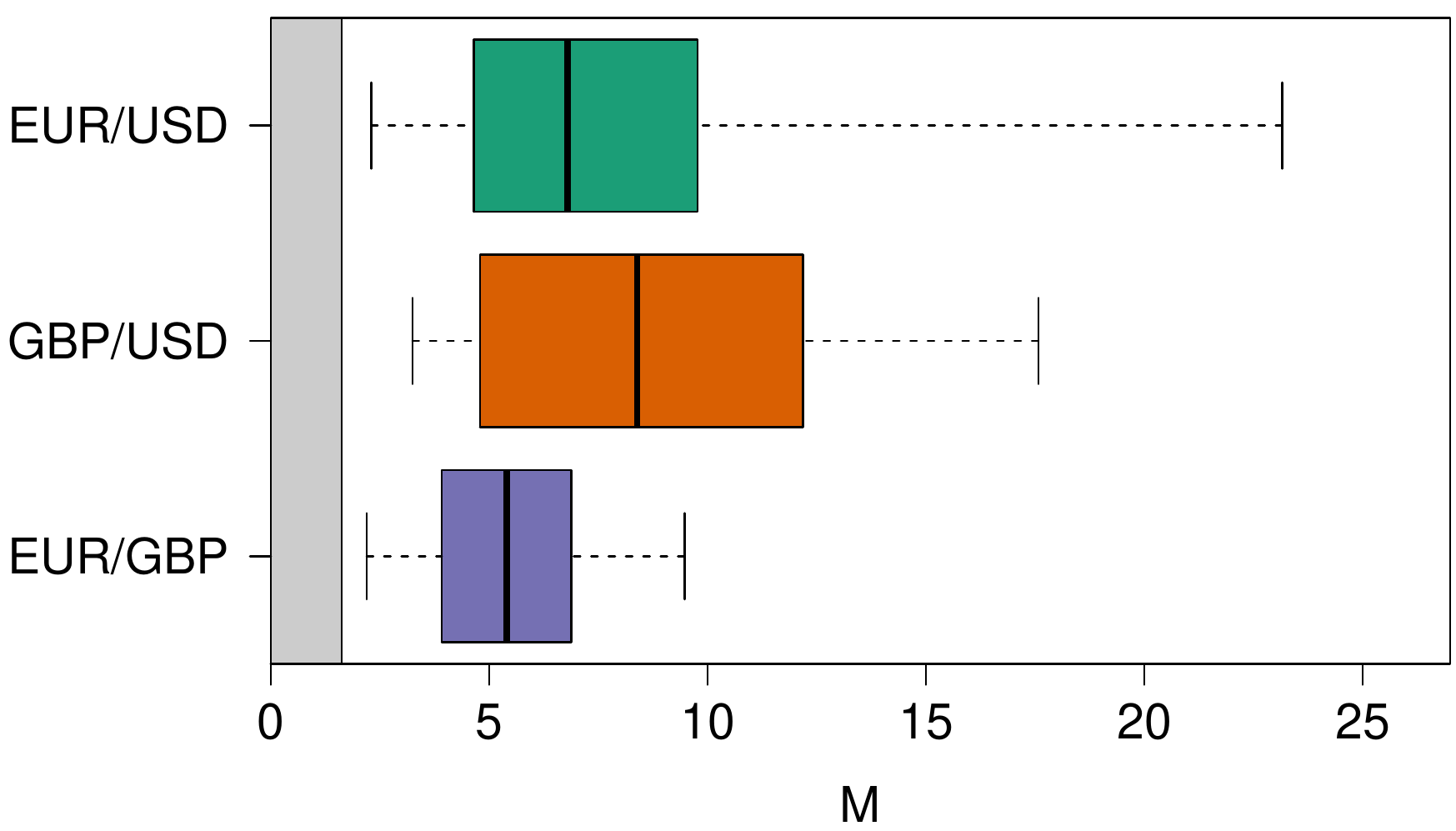}
\includegraphics[width=0.49\textwidth]{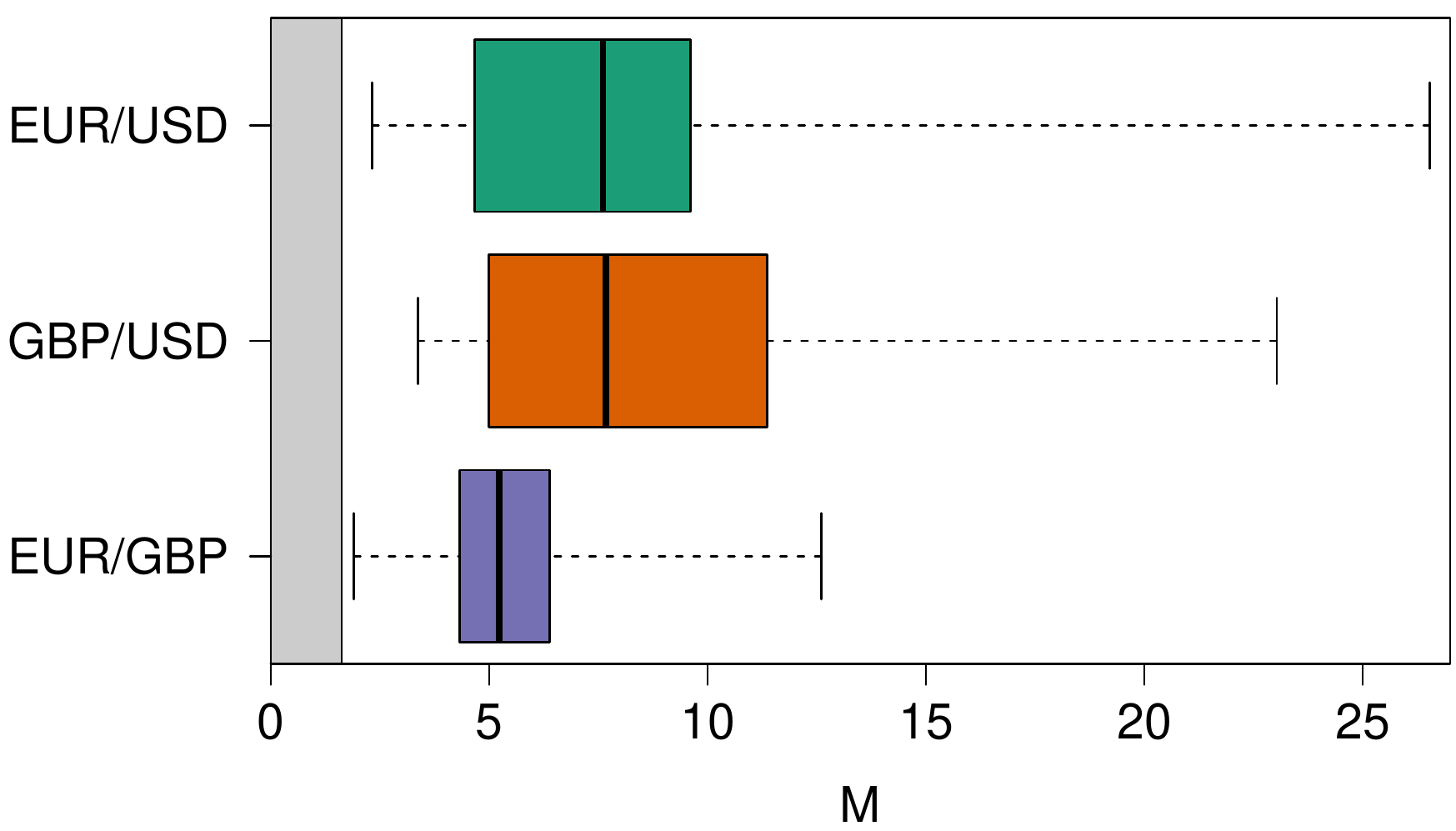}
\includegraphics[width=0.49\textwidth]{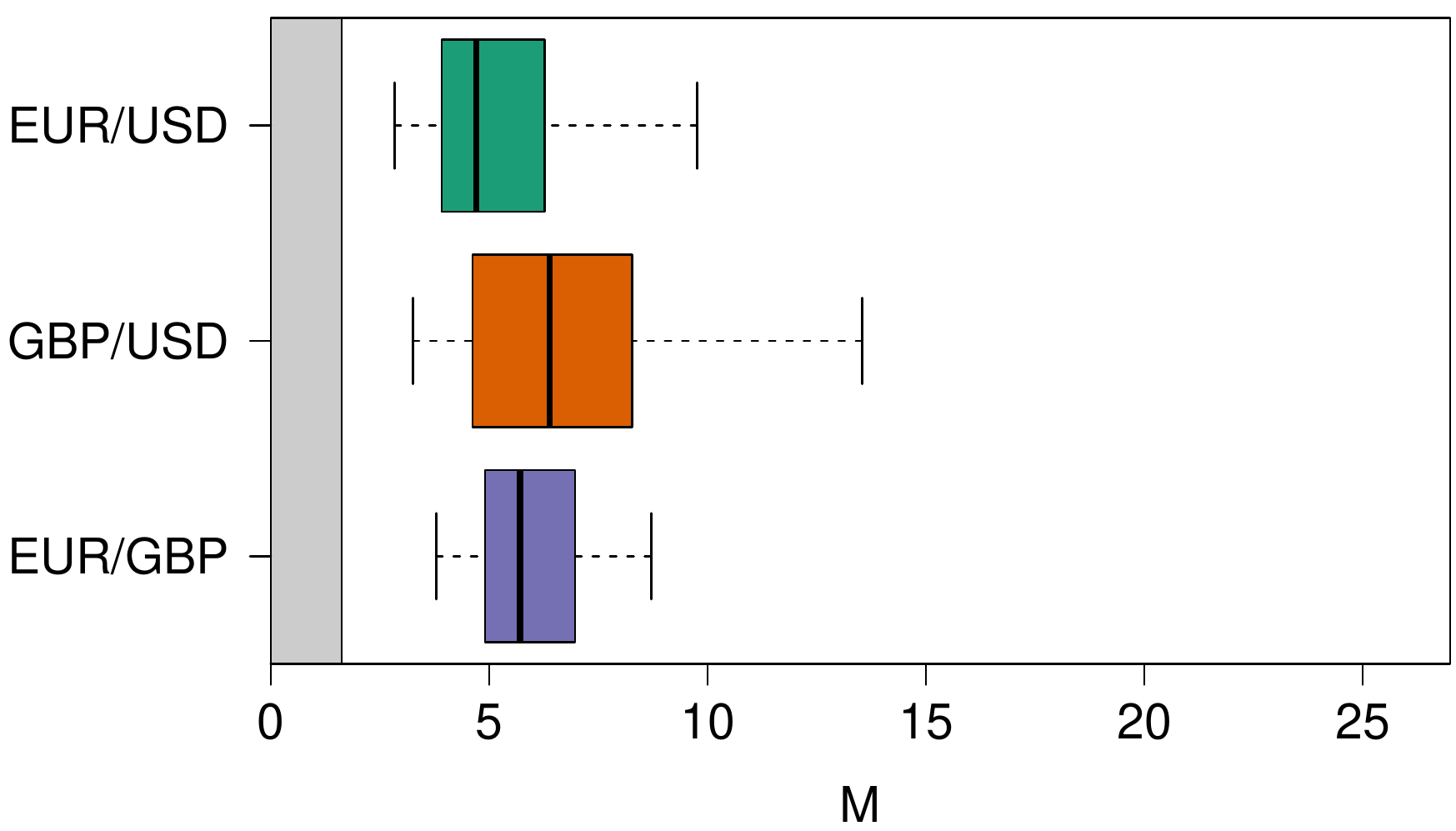}
\includegraphics[width=0.49\textwidth]{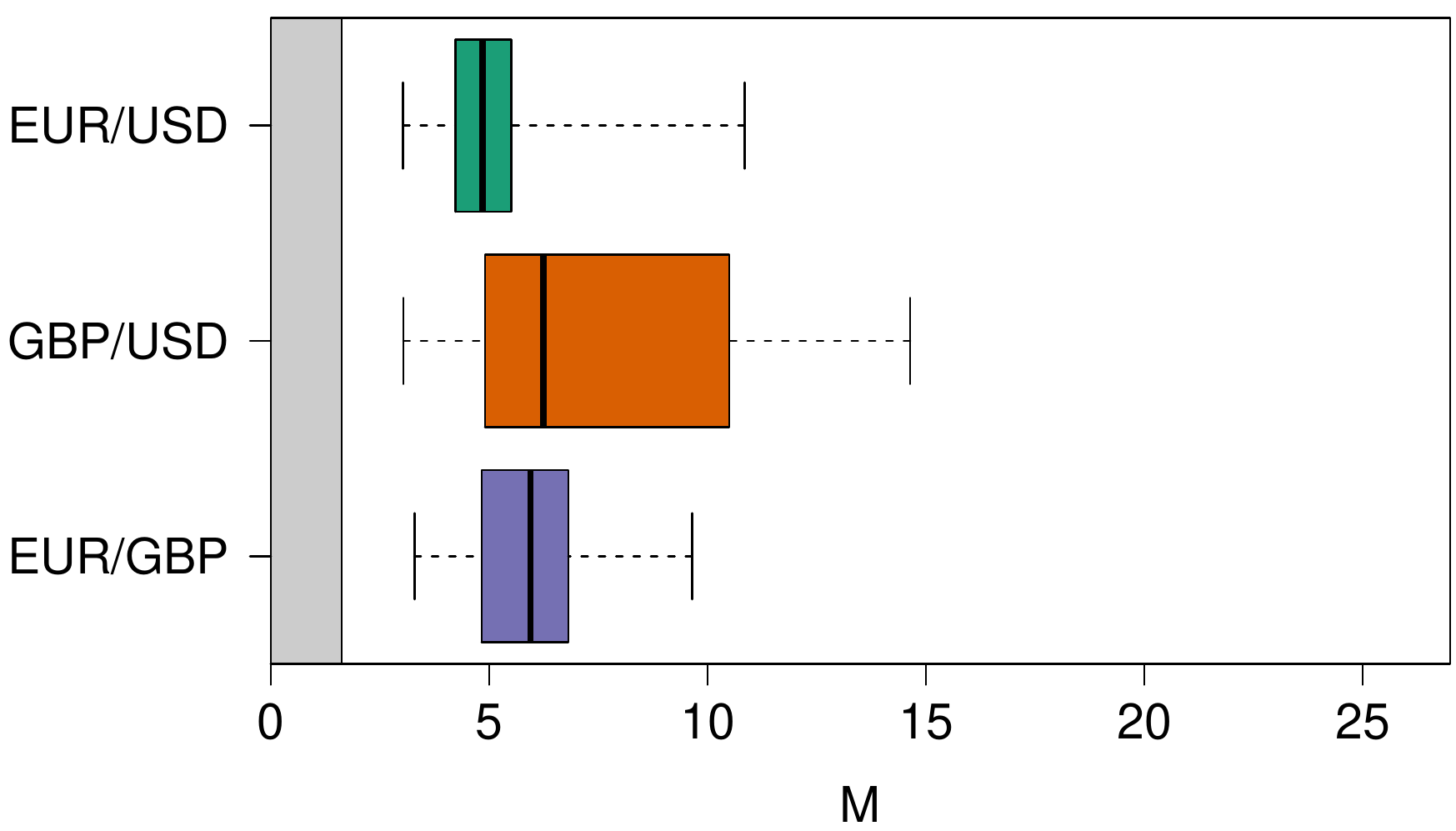}
\includegraphics[width=0.49\textwidth]{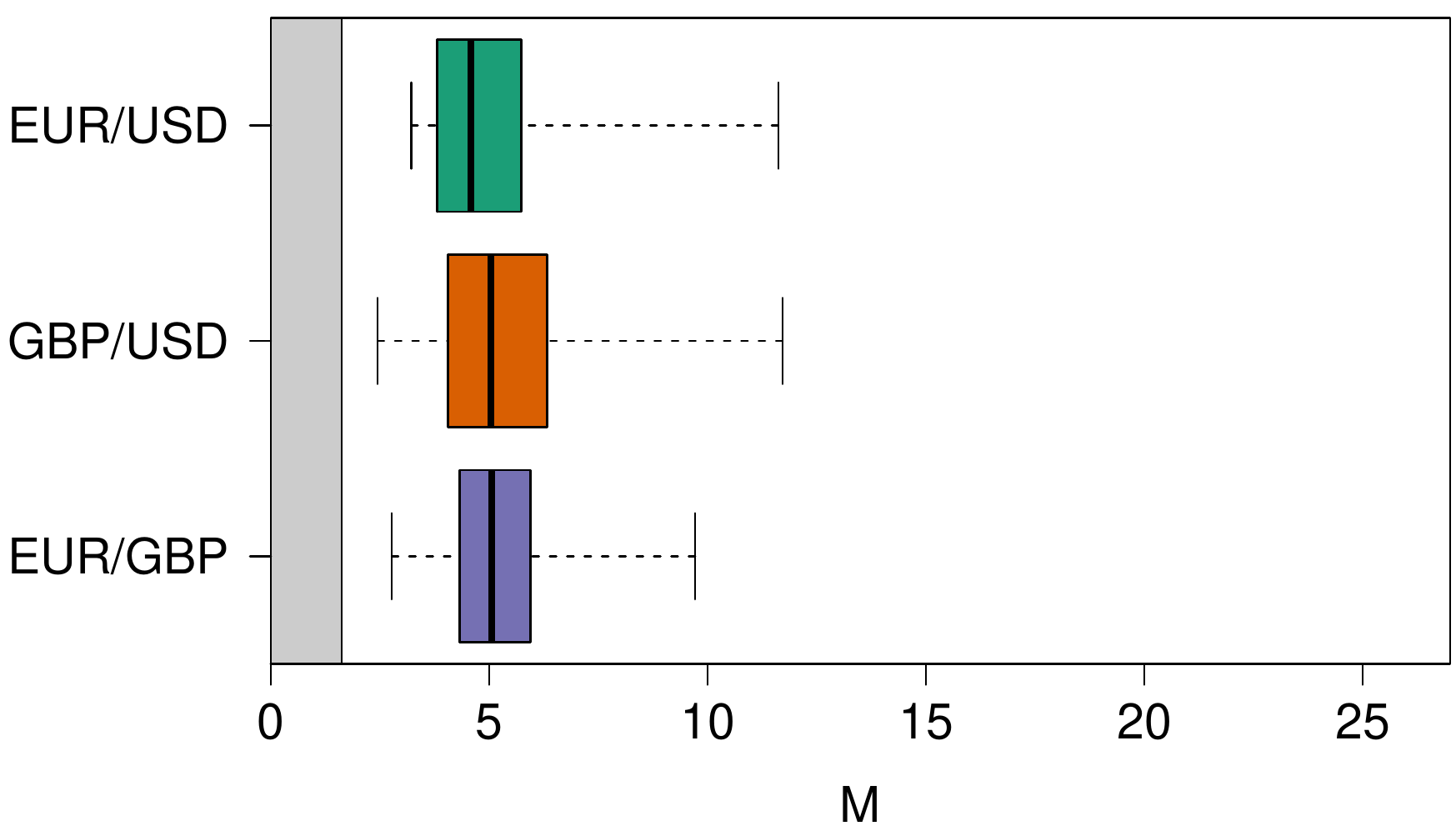}
\includegraphics[width=0.49\textwidth]{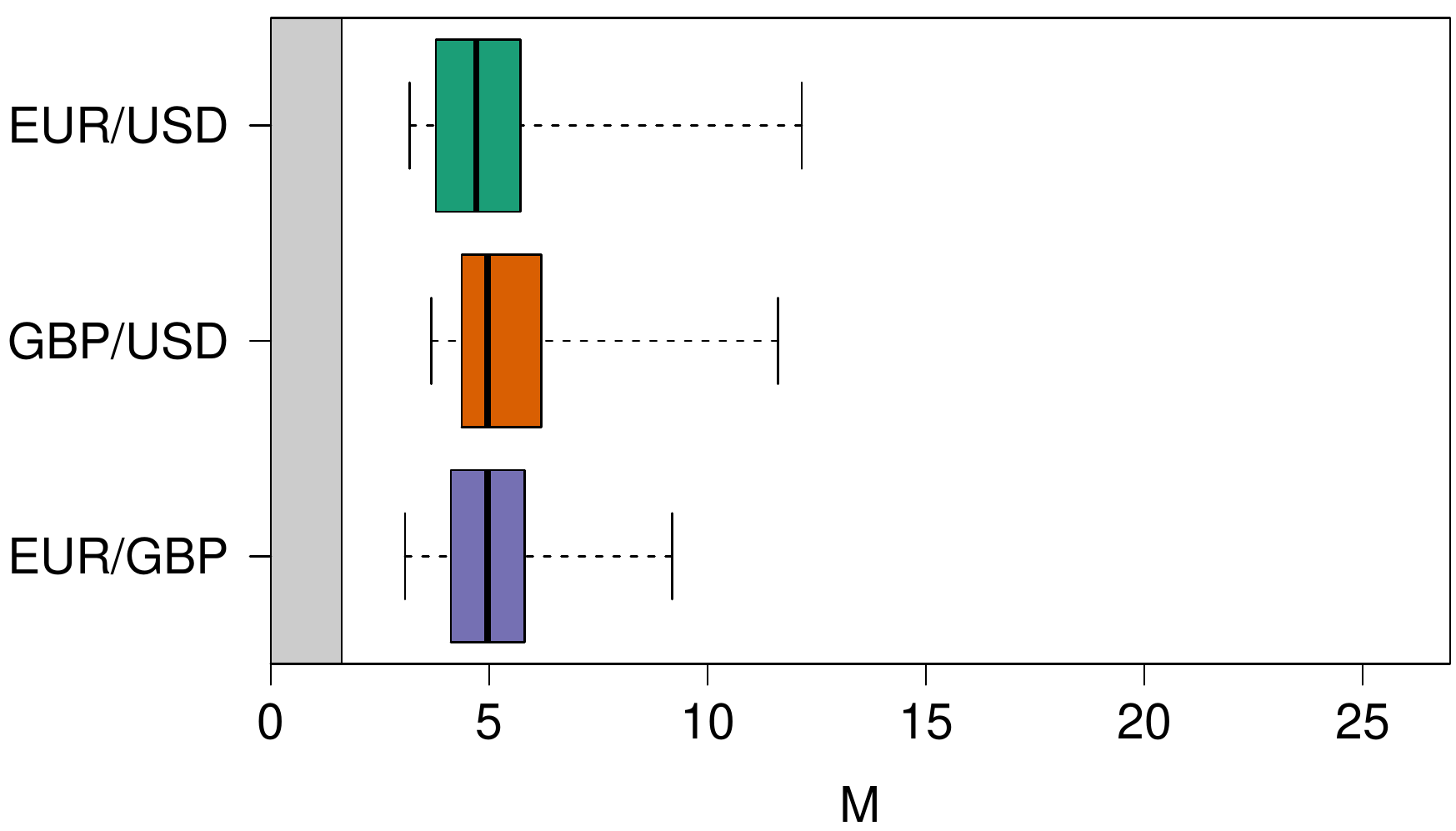}
\caption{Box plots of Berkes' change-point test statistic $M$ (see \ref{app:changepoint}) with (top row) 0 change points, (middle row) 1 change point, and (bottom row) 2 change points, for EUR/USD, GBP/USD, and EUR/GBP arrival-sign series. The left plots show the results for the intra-day series, and the right plots show the results for the cross-day series. The boxes indicate the lower quartile, median, and upper quartile of $M$, and the whiskers indicate the minimum and maximum of $M$, across all intra-day or cross-day series. In each case, we use Andrews' \cite{Andrews:1991heteroskedasticity} data-driven plug-in estimator $\hat{q}$ from Equation (\ref{eq:lmandrewsqhat}) to calculate $M$. The light grey shading indicates the critical region for Berkes' change-point test at the $1\%$ significance level. The results for the cross-day departure-sign series are qualitatively similar.}
\label{fig:CUSUMM012}
\end{figure}

To test the hypothesis of nonstationarity more formally, we also perform Berkes' change-point test (see \ref{app:changepoint}) on both the intra-day and cross-day order-flow series. We repeat the test for each of 0, 1, and 2 change points. We show box plots of our results in Figure \ref{fig:CUSUMM012}.

For each of the three currency pairs and for each of 0, 1, and 2 change-points, the test rejects the null hypothesis at the $1\%$ significance level in all intra-day and cross-day arrival-sign and departure-sign series. The results for the intra-day series are again very similar to those for the cross-day series. This result provides strong evidence against the hypothesis that the apparent long-memory properties of the series are actually artifacts caused by nonstationarities. We therefore strongly reject Axioglou and Skouras' \cite{Axioglou:2011markets} hypothesis that the apparent long memory in the cross-day series is mostly an artifact caused by structural breaks at the daily boundaries, in favour of the alternative hypothesis of true long memory that persists across daily boundaries. We return to this discussion in Section \ref{sec:lmdiscussion}.

\section{Discussion}\label{sec:lmdiscussion}

The results of our statistical tests strongly support the hypothesis that both the arrival-sign and departure-sign series on Hotspot FX exhibit long memory that persists over several thousands of events. For each of the three currency pairs, our calculations suggest that this long memory in order flow can be characterized by a Hurst exponent of $H \approx 0.7$. The variation in this result --- both across the different currency pairs and across the different days in our sample --- is small.

Our results raise several interesting points for discussion. As we noted in Section \ref{sec:litrev}, several authors have argued that order splitting provides a more plausible explanation for long memory in order flow than does herding. Our results provide further evidence to support this argument. Other empirical studies of long memory in order flow have studied LOBs in which all institutions observe all order flow from all others. On Hotspot FX, by contrast, institutions can only see the order flow that originates from institutions with whom they possess bilateral credit (see Section \ref{sec:data}). It is therefore reasonable to assume that herding effects are much weaker on Hotspot FX than they are on other platforms. Despite this important difference, our estimates of $H$ on Hotspot FX are very similar to those reported for other platforms (see Section \ref{sec:litrev}). We therefore argue that herding plays a minor role in the long-memory properties of order flow and that long-range autocorrelations are instead caused by order-splitting strategies (which are unaffected by the quasi-centralized nature of trade on Hotspot FX).

We turn next to the low-order negative autocorrelation that we observe in the sample ACFs (see Figure \ref{fig:ACFEU}). Although this negative autocorrelation is relatively weak, the effect is present for each day in our sample, and we therefore deem it to be a robust statistical property of the data. Interestingly, several other studies of long memory in order flow have not reported this behaviour and have instead reported positive autocorrelations at small lags. This raises the question of why our results differ. We believe that the answer to this question lies in the extremely high activity levels in the FX spot market. In a study of order flow on the LSE, T\'{o}th \emph{et al.} \cite{Toth:2012how} used broker identifier codes to trace which order flow originated from which brokerage. For limit order arrivals, they reported that orders originating from the same broker were positively correlated with each other, but that orders originating from different brokers were negatively correlated with each other. Because of the extremely large number of participants in the FX spot market, an arriving limit order is likely to be followed by many other arriving limit orders from other participants. According to T\'{o}th \emph{et al.}'s findings, this should generate short-range, negative autocorrelations in order flow. At larger lags, however, the strength of this effect diminishes and the long-range autocorrelation effects become more apparent. We conjecture that the short-range negative autocorrelations were not reported in some other studies because this effect is weaker (and, therefore, not observable) in markets with lower activity levels.

It is also interesting to consider why the cumulative-sum change-point estimator (see Equation \ref{eq:berkesk}) performs much worse at detecting the daily boundaries between intra-day order-sign series in the FX spot market than was reported by Axioglou and Skouras for the LSE \cite{Axioglou:2011markets}. We conjecture that the answer lies in the structural differences between the trading days in these two markets. Trading on the LSE commences at $08$:$00$:$00$ and ceases at $16$:$30$:$00$ each day \cite{LSEWebsiteSETS}. By contrast, trading in the FX spot market occurs 24 hours a day. Although we restrict our attention to the peak trading hours of $08$:$00$:$00$--$17$:$00$:$00$ GMT, the absence of a market-wide closing time has several important consequences for the way that traders act.

First, many financial institutions require that traders unwind their positions (i.e., rebalance their net daily holdings to $0$) before the end of each trading day \cite{Lyons:1995tests}. The LSE market closure at $16$:$30$:$00$ constitutes a hard deadline by which any traders who seek to unwind their positions must fulfill this goal, even if doing so requires them to trade at unfavourable prices. This may cause the statistical properties of order flow late in the trading day to differ substantially from those early in the trading day, and may therefore result in a structural break at the daily boundaries when concatenating data from different days. In the FX spot market, by contrast, there is no market-wide closing time.

Second, the absence of market opening and closing times in the FX spot market enables traders in different time zones to begin and end their trading days at different times. Hsieh and Kleidon \cite{Hsieh:1996bid} noted that many traders spend the early part of their trading day assessing the state of the market, the middle part of their trading day performing the majority of their trades, and the late part of their trading day resetting their net inventory to 0. In markets with specified opening and closing times (such as the LSE), all traders progress through this cycle simultaneously. In the FX spot market, by contrast, traders in different time zones can choose the length and timing of their trading days as they wish. The flux of traders from different time zones into and out of the FX spot market may cause the statistical properties of order flow to differ from those in markets where all traders' trading days are aligned by the market opening and closing times.

Third, Axioglou and Skouras \cite{Axioglou:2011markets} conjectured that most traders on the LSE reassess their trading strategy once per day, while markets are closed, then implement their chosen strategy throughout the next trading day. This is not a plausible description of the actions of traders in the FX spot market, which is always open.

Finally, we return to the question of whether the apparent long memory that we observe is actually an artifact caused by structural breaks. Although we test and strongly reject this hypothesis for zero, one, and two structural breaks, our results do not rule out the possibility of larger numbers of structural breaks. Indeed, it is plausible that intra-day order-sign series contain very large numbers of structural breaks, for reasons such as the arrivals or departures of institutions in the market or the release of macroeconomic news. From a modelling perspective, however, models that rely on large numbers of structural breaks that occur at unknown points in time have several important drawbacks, such as being difficult to estimate and of little use for forecasting (see \ref{app:changepoint}). Moreover, due to their extremely large numbers of parameters, such models suffer a considerable risk of overfitting. By contrast, long-memory models are parsimonious, easy to simulate, and useful for forecasting. We therefore regard long-memory approaches to be more useful than alternative approaches that rely on fitting large numbers of structural breaks to explain the observed autocorrelation properties of the order-sign series.

It is important to note that our results do not rule out the possibility that order-sign series contain \emph{both} long memory and structural breaks. Indeed, the co-existence of these effects could help to explain the differences between the behaviour observed in different markets. According to our findings, intra-day order flow in the FX spot market exhibits long memory. Any possible structural breaks appear to have little impact.  In equities markets, by contrast, the relatively low number of order arrivals each day makes precise empirical assessment of long memory much more difficult, and structural breaks appear to have a greater impact on the apparent long-memory properties of order flow. It is plausible that both structural breaks and true long memory could coexist in all financial markets and that the relative importance of these effects is different for different assets.

\section{Conclusions}\label{sec:lmconclusions}

In this paper, we have investigated the long memory of order flow in the FX spot market. Due to the extremely high levels of activity on the platform that we study, and in contrast to other empirical studies on this topic, we were able to investigate the long-memory properties of intra-day series without needing to aggregate data from different trading days. For each of the three currency pairs and on each of the trading days that we studied, we found that both arrival-sign and departure-series exhibit long memory, with a Hurst exponent of $H \approx 0.7$. We also uncovered a negative autocorrelation at shorter lags, which we conjectured was caused by the large number of participants in the FX spot market.

All of our results for data that crosses daily boundaries were similar to those for intra-day data, and we strongly rejected the hypothesis that the apparent long memory of order flow is an artifact caused by structural breaks. We therefore concluded that long memory is a robust statistical property of order flow on Hotspot FX that persists across daily boundaries. We also proposed several possible reasons why our findings differ from those reported by Axioglou and Skouras for the LSE \cite{Axioglou:2011markets}. Further empirical study of data from other markets will help to illuminate these issues further, and is therefore an important topic for future research.

Finally, we note that the existence of long memory in order flow raises an interesting question called the ``efficiency paradox'' \cite{Farmer:2006market}: how can return series remain unpredictable given that order flow exhibits long memory? To date, there are two main hypotheses. Some authors \cite{Bouchaud:2004fluctuations,Bouchaud:2006random} have argued that markets reside at a ``self-organized critical point'' in which liquidity takers cause long-range autocorrelations in order flow that exactly balance the long-range \emph{negative} autocorrelations caused by liquidity providers. Others \cite{Farmer:2006market,Lillo:2004long} have argued that predictability in order flow is offset by a negative correlation with available liquidity. At present, there is no clear consensus as to which approach best describes the temporal evolution of real markets, and further empirical and theoretical study of this question remains an important and exciting avenue for future research.

\appendix

\section{Long Memory}\label{app:lmtests}

In this appendix, we provide a detailed description of long-memory processes. For further discussion of these topics, see \cite{Beran:1994statistics}.

\subsection{Autocorrelation and Long Memory}

Recall from Section \ref{subsec:longmemory} that a second-order stationary time series $\left\{W_t\right\}=W_1,W_2,\ldots$ with ACF $\rho(k)$ (see Equation (\ref{eq:tsacf})) is said to exhibit long memory if\begin{equation}\lim_{N \rightarrow \infty}\sum_{k=-N}^N \left| \rho(k) \right|=\infty.\end{equation}

One way in which a time series can exhibit long memory is if there exists some constant $\alpha \in (0,1)$ such that $\rho(k)$ decays asymptotically as a power of $k$:\begin{equation}\label{eq:placf}\rho(k) \sim k^{-\alpha}L(k),\quad k \rightarrow \infty,\end{equation}where $L$ is a slowly varying function\footnote{A function $L$ is \emph{slowly varying} if $\lim_{k \rightarrow \infty} L(zk)/L(k) = 1$ for all $z>0$.} \cite{Cont:2005long, Lillo:2004long, Lillo:2005theory}. Smaller values of $\alpha$ correspond to slower decay of the long-range autocorrelations in $\left\{W_t\right\}$ \cite{Bouchaud:2009digest, Lillo:2004long}.

\subsection{The Rescaled-Range Statistic}\label{subsec:rescaledrange}

Let\begin{equation}\overline{W}_k=\frac{1}{k}\sum_{j=t+1}^{t+k}W_{j}.\end{equation}The \emph{rescaled-range statistic} \cite{Mandelbrot:1969computer2} is the ratio\begin{equation}\label{eq:lmadjrescaledrange}Q(t,k)=\frac{R(t,k)}{S(t,k)},\end{equation}where, for $t,k \in \mathbb{Z}_{>0}$ and for $i \in \left\{1,2,\ldots,k\right\}$,\begin{equation}\label{eq:lmrsR}R(t,k)=\max_{1 \leq i \leq k} \left[\sum_{j=t+1}^{t+i}\left(W_j-\overline{W}_k\right)\right]-\min_{1 \leq i \leq k} \left[\sum_{j=t+1}^{t+i}\left(W_j-\overline{W}_k\right)\right]\end{equation}and\begin{equation}\label{eq:lmrsS}S^2(t,k)=\frac{1}{k}\sum_{j=t+1}^{t+k}\left(W_j-\overline{W}_k\right)^2.\end{equation}The rescaled-range statistic $Q$ measures the range of partial sums of deviations of the time series $\left\{W_1,W_2,\ldots\right\}$ from its mean, rescaled by an estimate of its standard deviation \cite{Lo:1991long}.

\subsection{The Hurst Exponent}\label{subsec:hurst}

The following theorem by Mandelbrot \cite{Mandelbrot:1975limit} provides a relationship between a time series' long-range autocorrelations and its rescaled-range statistic.

\begin{mythe}If $\left\{W_t\right\}$ is a second-order stationary process such that $W_t^2$ is ergodic and $t^{-H}\sum_{i=1}^t W_i$ converges weakly to a fractional Brownian motion\footnote{A \emph{fractional Brownian motion} \cite{Mandelbrot:1968fractional} is a Gaussian process $B_H(t)$ with 0 drift that satisfies $B_H(0)=0$ and $\mathbb{E}\left[B_H(t)B_H(s)\right]=\frac{1}{2}\left(\left|t\right|^{2H}+\left|s\right|^{2H}-\left|t-s\right|^{2H}\right)$ for some $H\in(0,1)$.} with parameter $H$ as $t \rightarrow \infty$, then\begin{equation}\label{eq:lmmandelbrotresult}k^{-H}Q(t,k) \xrightarrow[]{d} \eta \text{ as }k \rightarrow \infty,\end{equation} where $\eta$ is a nondegenerate random variable and $\xrightarrow[]{d}$ denotes convergence in distribution.\end{mythe}The constant $H$ is called the \emph{Hurst exponent} of $\left\{W_t\right\}$ \cite{Beran:1994statistics,Hurst:1951long,Mandelbrot:1968noah, Mandelbrot:1969computer1,Mandelbrot:1969computer2,Mandelbrot:1969robustness}. A time series with a Hurst exponent of $H=1/2$ is a short-memory process.  For a long-memory process that satisfies the conditions of this theorem, $H$ is related to $\alpha$ in Equation (\ref{eq:placf}) by \begin{equation}\label{eq:hurstalpha}H=1-\frac{\alpha}{2}\end{equation}and to $\beta$ in Equation (\ref{eq:specdenslong}) by\begin{equation}\label{eq:hurstbeta}H=\frac{\beta+1}{2}.\end{equation}

\section{Empirical Assessment of Long Memory}\label{sec:lmtests}

In many empirical situations, it is common to observe only a single, finite-length realization $\left\{w_1,w_2,\ldots,w_N\right\}$ of $\left\{W_t\right\}$.  If the statistical properties of $\left\{W_t\right\}$ are unknown, then estimating the long-memory properties of $\left\{W_t\right\}$ from $\left\{w_1,w_2,\ldots,w_N\right\}$ entails considerable challenges \cite{Beran:1992statistical, Beran:1994statistics, Mandelbrot:1969some}. Most empirical studies employ heuristic methods for this task. The performance of such techniques on empirically observed series varies considerably, so it is common for empirical studies to evaluate the output of several heuristic methods rather than relying on a single estimator. In this appendix, we provide a detailed description of the techniques that we use throughout the paper. For further discussion and comparisons of these techniques, see \cite{Taqqu:1995estimators}.

\subsection{Sample Autocorrelation Function}\label{subsec:sampleacf}

For an empirically observed time series $\left\{w_1,w_2,\ldots,w_N\right\}$, let\begin{equation}\overline{w} = \frac{1}{N}\sum_{i=1}^N w_i\end{equation}denote the sample mean, let\begin{equation}\label{eq:sampleautocov}\hat{\gamma}(k)=\frac{1}{N}\sum_{i=1}^{N-\left|k\right|}\left( w_{i+\left|k\right|}-\overline{w}\right)\left(w_{i}-\overline{w}\right)\end{equation}denote the sample autocovariance function, and let\begin{equation}\label{eq:sampleacf}\hat{\rho}(k)=\frac{\hat{\gamma}(k)}{\hat{\gamma}(0)}\end{equation}denote the sample ACF. It is very difficult to estimate the large-$k$ decay of $\rho(k)$ from $\hat{\rho}$, so direct estimation of the long-memory properties of $\left\{W_t\right\}$ from $\hat{\gamma}$ often produces very poor results \cite{Lillo:2004long}.

\subsection{Rescaled-Range Plots}\label{subsec:lmpox}

For a given \emph{block number} $B\in \mathbb{N}$, let\begin{equation}\label{eq:appoints}G(k)=\left\{t = \frac{N(i-1)}{B}+1 \vert i=1,\ldots,B; \ t+k\leq N \right\}.\end{equation}A \emph{rescaled-range plot} (also known as a \emph{pox plot}) \cite{Mandelbrot:1968noah, Mandelbrot:1969computer2, Teverovsky:1999critical} is a plot of\begin{equation}\label{eq:poxmean}\overline{R}(k)=\frac{1}{\left|G(k)\right|}\sum_{t \in G(k)} Q(t,k)\end{equation}versus $k$ on doubly logarithmic axes, where $\left|G(k)\right|$ denotes the number of elements in $G(k)$. The slope of a rescaled-range plot for large values of $k$ provides a rough estimate for the Hurst exponent $H$ \cite{Mandelbrot:1969computer2}.

\subsection{Lo's Modified Rescaled-Range Statistic}\label{subsec:lmlo}

Lo \cite{Lo:1991long} noted that if a time series $\left\{W_t\right\}$ is subject to short-range autocorrelations, then the denominator $S(t,k)$ of the rescaled-range statistic $Q(t,k)$ is not a consistent estimator for the standard deviation of $\left\{W_t\right\}$.  Therefore, an important difficulty in using the rescaled-range statistic to assess the long-memory properties of an empirically observed time series $\left\{w_1,w_2,\ldots,w_N\right\}$ is that the finite-sample properties of $Q$ are not invariant to short-range dependence. To address this problem, Lo proposed replacing the denominator of $Q$ with the Newey--West\footnote{For a suitable choice of $q$, Newey and West \cite{Newey:1987simple} showed that $\hat{\sigma}(q)$ is a consistent estimator for the standard deviation of $\left\{W_t\right\}$, even if $\left\{W_t\right\}$ is subject to short-range autocorrelations.} estimator \cite{Newey:1987simple}, which discounts short-range dependence in $\left\{W_t\right\}$ up to a specified lag $q<N$. The parameter $q$ is called the \emph{bandwidth parameter}. Lo's modified rescaled-range statistic \cite{Lo:1991long} is\begin{equation}\label{eq:loQ}\tilde{Q}(q)=\frac{R(1,N)}{\hat{\sigma}(q)},\end{equation}where\begin{equation}\label{eq:lmrsSigma}\hat{\sigma}^2(q)=\left\{ \begin{array}{ll}S^2(1,N),&\text{ if }q=0,\\
S^2(1,N)+2\sum_{i=1}^q \left( 1-\frac{i}{q+1} \right) \hat{\gamma}(i),&\text{ otherwise.}\end{array}\right.\end{equation}

Given $\tilde{Q}(q)$, the statistic\begin{equation}\label{eq:Vk}V(q)=\frac{\tilde{Q}(q)}{\sqrt{N}}\end{equation}can be used as a test statistic for the hypothesis test\begin{align*}H_0&:\left\{W_t\right\} \text{ is a short-memory process,}\\
H_1&:\left\{W_t\right\} \text{ is a long-memory process.}\end{align*}This hypothesis test is called \emph{Lo's modified rescaled-range test} \cite{Lo:1991long}.  In the limit $N \rightarrow \infty$, the asymptotic critical region for the test at the $5\%$ significance level is approximately $\left[0.809,1.862\right]$.

Teverovsky \emph{et al.} \cite{Teverovsky:1999critical} remarked that although Lo's modified rescaled-range statistic is a significant improvement over the original rescaled-range statistic, Lo's test can fail to reject $H_0$ for some time series with long memory. Moreover, they noted that both the size and the power of the test depend on $q$. The optimal choice of $q$ for $\tilde{Q}(q)$ and $V(q)$ depends on the behaviour of the spectral density $f$ of $\left\{W_t\right\}$ \cite{Andrews:1991heteroskedasticity}.  If $f$ is unknown (as is usually the case for empirically observed series), then there is no universal rule via which to choose $q$. In empirical applications, it is therefore customary to calculate $\tilde{Q}(q)$ and $V(q)$ using several different choices of $q$ and/or to calculate a so-called \emph{plug-in estimator} $\hat{q}$ \cite{Andrews:1991heteroskedasticity,Axioglou:2011markets,Lo:1991long} by assuming that $f$ is equal to the spectral density of a specified parametric process. Andrews \cite{Andrews:1991heteroskedasticity} derived plug-in estimators for several different parametric processes (including autoregressive, moving-average, and ARMA models).

For our calculations, we use the plug-in estimator for an AR(1) process\begin{equation}W_{t}=\phi W_{t-1} + \varepsilon_{t},\end{equation}where $\phi \in \mathbb{R}$ is the autocorrelation parameter and $\varepsilon_{t}$ is uncorrelated Gaussian noise. This estimator is given by \cite{Andrews:1991heteroskedasticity,Lo:1991long}\begin{equation}\label{eq:lmandrewsqhat}\hat{q}=\left\lfloor \left(\frac{3N}{2}\right)^{1/3}\left(\frac{2\hat{\phi}}{1-\hat{\phi}^2}\right)^{2/3}\right\rfloor,\end{equation}where $\hat{\phi}$ is the maximum-likelihood estimate of $\phi$ given $\left\{w_1,w_2,\ldots,w_N\right\}$ and $\left\lfloor x\right\rfloor$ denotes the greatest integer less than or equal to $x$.

\subsection{Detrended Fluctuation Analysis}\label{subsec:lmdfa}

\emph{Detrended fluctuation analysis (DFA)} \cite{Peng:1994mosaic} is a technique for estimating the Hurst exponent from an empirically observed series $\left\{w_1,w_2,\ldots,w_N\right\}$. Let\begin{equation}\label{eq:lmwstar}w_i^*=\sum_{j=1}^i w_i,\ i=1,2,\ldots,N.\end{equation}For a given window length $m \in \mathbb{N}$ such that $m \leq N$, divide $\left\{w_1^*,w_2^*,\ldots,w_N^*\right\}$ into non-overlapping windows of length $m$.  For each $j\in \left\{1,2,\ldots,\left\lfloor{N/m}\right\rfloor\right\}$, label the data points in window $j$ as $y_{1,j},y_{2,j},\ldots ,y_{m,j}$, perform an ordinary least-squares regression to fit a straight line to the $m$ data points in the window, and let $\hat{y}_{i,j}$ denote the value of the regression line at the point $y_{i,j}$ for each $i\in \left\{1,2,\ldots,m\right\}$.  For each window, calculate the \emph{detrended standard deviation}\begin{equation}\label{eq:dfawindow}F_j(m)=\sqrt{\frac{1}{m}\sum_{i=1}^m\left(y_{i,j}-\hat{y}_{i,j}\right)^2},\end{equation}and then calculate the \emph{length-$m$ mean detrended standard deviation}\begin{equation}\label{eq:dfaoverall}F(m)=\frac{1}{\left\lfloor{N/m}\right\rfloor}\sum_{j=1}^{\left\lfloor{N/m}\right\rfloor}F_j(m).\end{equation}Repeat this process for several logarithmically spaced choices of window length $m$, and plot $F(m)$ versus $m$ using doubly logarithmic axes.  Identify the smallest value $m_{\min}$ such that the plot is approximately straight for all $m\geq m_{\min}$.  The DFA estimate of $H$ is given by the slope of the best-fit line for $m\geq m_{\min}$.

\subsection{Log-Periodogram Regression}\label{subsec:lmlp}

The long-memory properties of $\left\{W_t\right\}$ can also be characterized by the behaviour of its spectral density \cite{Beran:1994statistics}\begin{equation}\label{eq:lmspectraldensity}f(\lambda) = \frac{1}{2\pi} \sum_{k=-\infty}^{\infty} \gamma(k) e^{-ik\lambda}\end{equation}in the limit $\lambda \rightarrow 0$. If there exists a constant $l \in \mathbb{R}$ such that\begin{equation}\label{eq:specdensshort}f(\lambda)\rightarrow l\text{ as }\lambda\rightarrow 0,\end{equation}then $\left\{W_t\right\}$ is a short-memory process. If, by contrast, there exists a constant $\beta \in \left(0,1\right)$ such that\begin{equation}\label{eq:specdenslong}f(\lambda)\sim \lambda^{-\beta} \text{ as }\lambda\rightarrow 0,\end{equation}then $\left\{W_t\right\}$ is a long-memory process.  Larger values of $\beta$ correspond to slower decay of the long-range autocorrelations in $\left\{W_t\right\}$ \cite{Beran:1994statistics}.

Estimating the behaviour of $f(\lambda)$ close to $0$ provides an alternative approach to estimating the Hurst exponent $H$.  Let\begin{equation}\label{eq:periodogramlambda}\lambda_{j,N}=\frac{2 \pi j}{N}, \qquad j \in \left\{1,2,\ldots,\left\lfloor\frac{N-1}{2}\right\rfloor\right\}\end{equation}denote the \emph{Fourier frequencies} of $\left\{w_1,w_2,\ldots,w_N\right\}$, and let\begin{equation}\label{eq:lmnewperiodogram}I(\lambda_{j,N})=\frac{1}{2\pi N} \left| \sum_{t=1}^{N}\left(w_t-\overline{w}\right)e^{-it\lambda_{j,N}}\right|^2\end{equation}denote the corresponding \emph{periodogram}. The slope of an ordinary least-squares regression of $\log\left(I(\lambda_{j,N})\right)$ onto $\log\left(\lambda_{j,N}\right)$ for small $\lambda_{j,N}$ is an estimator for $-\beta$, and it is therefore an estimator for $H$ \cite{Beran:1994statistics,Lillo:2004long,Taqqu:1995estimators}.

Despite the attractiveness of its computational simplicity, log-periodogram regression suffers from a substantial practical drawback \cite{Beran:1994statistics}: there is no universal rule for choosing the number $c$ of Fourier frequencies with which to perform the regression.  The scaling in Equations (\ref{eq:specdensshort}) and (\ref{eq:specdenslong}) only holds for $\lambda\rightarrow 0$, so choosing an overly large $c$ leads to large bias, but choosing an overly small $c$ leads to high variance. Robinson \cite{Robinson:1994rates} derived an expression for the optimal choice of $c$ to minimize the mean squared error of the estimated cumulative spectral distribution function, but the optimal choice turns out to depend on the unknown value of $H$.  Therefore, Robinson's expression does not provide a method for choosing $c$ for an empirically observed time series.  Instead, most empirical studies use one of two rules of thumb: $c=\sqrt{N}$ \cite{Geweke:1983estimation} or $c=0.1\times(N/2)$ \cite{Taqqu:1995estimators}. Despite their widespread use, neither of these rules are based on rigorous derivation or optimization.

\section{Long Memory versus Nonstationarity}\label{app:changepoint}

A key difficulty with assessing the long-memory properties of empirically observed time series is that many estimation techniques can produce similar output for a nonstationary series (e.g., a series with a monotonic trend \cite{Bhattacharya:1983hurst} or change in mean \cite{Giraitis:2001testing,Granger:2004occasional}) as they do for a stationary series with long memory \cite{Rea:2009estimators,Taqqu:1995estimators,Xu:2005quantifying}. Several authors have thus argued that the apparent long memory reported by empirical studies of financial time series is an artifact caused by nonstationarities \cite{Axioglou:2011markets,Berkes:2006discriminating,Liu:2000modeling,Mikosch:2002long}.


Disentangling the statistical properties of a time series with long memory and a time series with nonstationarities is a difficult task, and the choice of whether to model such time series using a long-memory model or a nonstationary model often depends on the desired application. Long-memory models are parsimonious, straightforward to simulate, and there exist many techniques that require only mild assumptions to estimate their parameters from data \cite{Beran:1994statistics}. Nonstationary models can illuminate important features of an underlying series (such as the locations and frequency of structural breaks) that are not addressed by long-memory models, but they typically require the inclusion of either a large number of parameters (which can lead to over-fitting) or latent parameters (which can be difficult to estimate from data).

Many standard tests for nonstationarities in an empirically observed series have low power in the presence of long memory \cite{Diebold:1991power,Hassler:1994power}, and there is no universal test that is able to determine whether the apparent long-memory properties of an empirically observed time series are an artifact caused by some unknown form of nonstationarity \cite{Granger:2004occasional}. However, Berkes \emph{et al.} \cite{Berkes:2006discriminating} developed a hypothesis test that can help to distinguish between long memory and a specific type of nonstationarity that they called \emph{structural breaks}.

\begin{mydef}The time series $\left\{Z_t\right\}$ is a \emph{short-range dependent series with a structural break at time $r^*$} if there exists a real-valued, second-order stationary, short-memory process $\left\{\zeta_t\right\}$ and a constant $\mu^* \neq 0$ such that\begin{equation}\label{eq:strucbreak2}Z_t = \left\{\begin{array}{ll}
\zeta_t,& t\leq r^*, \\
\mu^* + \zeta_t,& t> r^*.
\end{array}\right.\end{equation}\end{mydef}

Given a finite-length empirical observation $\left\{z_1,z_2,\ldots,z_N\right\}$ of a time series $\left\{Z_t\right\}$, \emph{Berkes' change-point test} is the hypothesis test\begin{align*}H_0&:\left\{Z_t\right\} \text{ is a short-memory process with one structural break,\footnote{For brevity, we restrict our discussion to Berkes' change-point test for a single structural break. However, this framework can be extended to test the null hypothesis of any specified number of structural breaks against the alternative hypothesis of true long memory. For full details, see \cite{Berkes:2006discriminating}.}}\\
H_1&:\left\{Z_t\right\} \text{ is a long-memory process.}\end{align*}Berkes' test uses the so-called \emph{cumulative-sum change-point estimator} \cite{Berkes:2006discriminating}\begin{equation}\label{eq:berkesk}\hat{r}^*=\min \left\{ r : \max_{1\leq j \leq N} \left|\sum_{i=1}^j z_i - \frac{j}{N} \sum_{i=1}^N z_i \right| = \left| \sum_{i=1}^{r} z_i-\frac{r}{N} \sum_{i=1}^N z_i \right| \right\}.\end{equation}For a given $q<N$, let\begin{align*}T^{(1)} &= \frac{1}{\hat{\sigma}_{(1)}^2(q) \sqrt{\hat{r}^*}}\max_{1 \leq i \leq \hat{r}^*}\left|\sum_{j=1}^i z_j-\frac{i}{\hat{r}^*}\sum_{j=1}^{\hat{r}^*}z_j\right|,\\
T^{(2)} &= \frac{1}{\hat{\sigma}_{(2)}^2(q) \sqrt{N-\hat{r}^*}}\max_{\hat{r}^* \leq i \leq N}\left|\sum_{j=\hat{r}^*}^N z_j-\frac{i-\hat{r}^*}{N-\hat{r}^*}\sum_{j=\hat{r}^*}^{N}z_j\right|,\end{align*}where $\hat{\sigma}_{(1)}(q)$ and $\hat{\sigma}_{(2)}(q)$ are the values of $\hat{\sigma}(q)$ from Equation (\ref{eq:lmrsSigma}) for the $\left\{z_1,z_2,\ldots,z_{\hat{r}^*}\right\}$ and $\left\{z_{\hat{r}^*+1},z_{\hat{r}^*+2},\ldots,z_N\right\}$ series, respectively.  The test statistic for Berkes' test is\begin{equation}\label{eq:berkesM}M=\max\left\{T^{(1)},T^{(2)}\right\}.\end{equation}In the limit $N \rightarrow \infty$, the asymptotic critical value of $M$ at the $1\%$ significance level is about $1.72$ \cite{Berkes:2006discriminating}.

\section*{Acknowledgements}We thank Julius Bonart, Jean-Philippe Bouchaud, Rama Cont, J. Doyne Farmer, Austin Gerig, Ben Hambly, and Gabriele La Spada for useful discussions. We thank Hotspot FX for providing the data for this project. We also thank Terry Lyons, Rich Plummer-Powell, and Justin Sharp for technical support. MDG and SH thank the Oxford-Man Institute of Quantitative Finance, and MDG thanks EPSRC and the James S. McDonnell Foundation for supporting this research.

\bibliography{dphilLOBbib}

\end{document}